\documentclass[reprint,amsmath,amssymb,aps,nofootinbib,superscriptaddress]{revtex4-1}
\usepackage{booktabs}

\usepackage[pdftex,colorlinks]{hyperref}
\usepackage[dvipsnames]{xcolor}
\definecolor{phthaloblue}{rgb}{0.0, 0.06, 0.54}

\hypersetup{
    colorlinks=true,
    linkcolor=magenta,
    citecolor=blue,
    filecolor=MidnightBlue,
    urlcolor=blue,
    }
\usepackage[pdftex]{graphicx}
\usepackage{bm}
\usepackage{eucal}  
\usepackage{mathrsfs}
\usepackage{cancel}
\usepackage{here}
\usepackage{comment,braket}
\usepackage{epsf}
\usepackage{xcolor} 
\usepackage{amsmath}
\usepackage{graphics}
\usepackage{amsfonts}
\usepackage{amssymb}
\usepackage{latexsym}
\usepackage{color}
\usepackage{natbib}
\usepackage[normalem]{ulem}
\input{colordvi.tex}
\usepackage{feynmp}
\usepackage[utf8]{inputenc}
\DeclareUnicodeCharacter{2212}{-}
\usepackage{orcidlink}
\usepackage{multirow}
\usepackage{subcaption}
\newcommand{\beq}{\begin{equation}}
\newcommand{\eeq}{\end{equation}}
\usepackage{bm}

\newcounter{notecountRJ}[section]

\def\lsim{\mathrel{\raise.3ex\hbox{$<$\kern-.75em\lower1ex\hbox{$\sim$}}}}
\def\gsim{\mathrel{\raise.3ex\hbox{$>$\kern-.75em\lower1ex\hbox{$\sim$}}}}

\usepackage{booktabs}
\usepackage{orcidlink}

\begin{document}

\title{Constraints on High-Frequency Gravitational Waves from Graviton–Photon Conversion in the M87 Galaxy}

 \author{Aman Gupta \orcidlink{0000-0002-7247-2424}}
 \email{amann16.iitr@gmail.com}
 \affiliation{School of Physical Sciences, Indian Association for the Cultivation of Science,\\ 2A \& 2B Raja S.C. Mullick Road, Kolkata-700032, India}
 \affiliation{Institute of High Energy Physics, Chinese Academy of Sciences, Beijing 100049, China}
 \affiliation{Spallation Neutron Source Science Center, Dongguan, Guangdong, 523803, China}

 \author{Pratik Majumdar \orcidlink{0000-0002-5481-5040}}
 \email{pratik.majumdar@saha.ac.in}
 \affiliation{Saha Institute of Nuclear Physics, 1/AF, Bidhannagar, Kolkata 700064, India} 
 \affiliation{Homi  Bhabha  National  Institute,  Anushakti  Nagar,  Mumbai  400094,  India}

 \author{Sourov Roy \orcidlink{0000-0002-1015-3241}}
 \email{tpsr@iacs.res.in}
 \affiliation{School of Physical Sciences, Indian Association for the Cultivation of Science,\\ 2A \& 2B Raja S.C. Mullick Road, Kolkata-700032, India}

 \author{Pratick Sarkar \orcidlink{0009-0000-8160-0734}}
 \email{spsps2523@iacs.res.in}
 \affiliation{School of Physical Sciences, Indian Association for the Cultivation of Science,\\ 2A \& 2B Raja S.C. Mullick Road, Kolkata-700032, India}

\begin{abstract}

High-frequency gravitational waves, particularly in the range $f \gtrsim 10^{10}~\mathrm{Hz}$, represent a compelling probe of physics beyond the Standard Model. Due to the absence of direct detection methods in this frequency regime, alternative strategies may be pursued. One promising approach involves the conversion of gravitons into photons in the presence of magnetic fields, a process known as the inverse Gertsenshtein effect. In this study, we explore such graviton-to-photon conversions occurring within the magnetic field environment of the M87 galaxy, utilizing realistic models for the galactic magnetic field and plasma density structure.
We use 
the broadband electromagnetic spectrum of M87, ranging from millimeter to TeV gamma rays, to search for hidden contributions from graviton-photon conversions. In the well-constrained frequency range $10^{10}$–$10^{27}~\mathrm{Hz}$, the lack of excess emission allows us to place improved bounds on the gravitational wave strain amplitude $h_c$ or on spectral energy density $\Omega_{\mathrm{gw}} h^2$.
We find that our results from M87 yield substantially stronger constraints compared to existing bounds derived from Milky Way magnetic field considerations, with improvements ranging from one to five orders of magnitude depending on the frequency band, thereby enhancing the prospects for probing high-frequency gravitational wave backgrounds through indirect electromagnetic signatures.

\end{abstract}

\maketitle

\section{Introduction}

For many decades, our understanding of the Universe has primarily relied on electromagnetic observations, spanning more than twenty orders of magnitude in frequency—from radio waves to high-energy gamma rays. 
Gravitational Waves (GWs) provide a fundamentally different probe of the cosmos, as they travel across cosmological distances largely unaffected by intervening matter. Compact binary mergers have now been directly detected through GWs in the frequency band of a few Hz to kHz by the LIGO–Virgo–KAGRA collaboration~\cite{LIGOScientific:2018mvr,LIGOScientific:2020ibl,KAGRA:2021vkt}. More recently, pulsar timing arrays (PTAs) have reported evidence for a stochastic gravitational wave background (SGWB) in the nanohertz (nHz) regime~\cite{NANOGrav:2023gor,EPTA:2023fyk,IPTA:2023ero,Reardon:2023gzh}

In contrast, high-frequency GWs, with frequencies $f \gtrsim \rm MHz$, are not expected to originate from conventional astrophysical sources such as compact binary mergers or stellar core collapses~\cite{Rosado:2011kv,Sesana:2016ljz,Lamberts:2019nyk,Robson:2018ifk,KAGRA:2021kbb,Aggarwal:2020olq,Babak:2023lro}. Instead, their detection could point toward exotic physics beyond the Standard Model. Various early universe phenomena are predicted to generate a high-frequency GWB, including inflation~\cite{Grishchuk:1974ny,Starobinsky:1979ty,Rubakov:1982df,Kim:2004rp,Peloso:2015dsa,Bartolo:2016ami,Domcke:2016bkh,Garcia-Bellido:2016dkw,Vagnozzi:2022qmc}, preheating and reheating~\cite{Figueroa:2017vfa,Kanemura:2023pnv}, first-order phase transitions~\cite{Caprini:2015zlo,Caprini:2018mtu,Caprini:2019egz,Hindmarsh:2020hop,Gouttenoire:2022gwi,Athron:2023xlk}, topological defects~\cite{Vilenkin:2000jqa, Blanco-Pillado:2017oxo, Auclair:2019wcv, Gouttenoire:2019kij,Servant:2023tua,Lazarides:2024niy,Maji:2026nkz}, black hole superradiance~\cite{Brito:2015oca} and primordial black holes~\cite{Anantua:2008am,Dolgov:2011cq,Fujita:2014hha,Dong:2015yjs,Herman:2020wao,Franciolini:2022htd,Gehrman:2022imk,Gehrman:2023esa}.

Despite their theoretical importance, the detection of high-frequency GWs remains far beyond the reach of current gravitational wave observatories such as LIGO, VIRGO, and KAGRA~\cite{LIGOScientific:2014qfs,LIGOScientific:2019vic}, as well as proposed future detectors such as LISA~\cite{LISA:2024hlh}, the Einstein Telescope~\cite{Hild:2010id, Punturo:2010zz}, and Cosmic Explorer~\cite{LIGOScientific:2016wof}. This challenge necessitates the development of alternative, indirect detection strategies.

To outmanoeuvre the lack of dedicated high frequency detectors, many indirect searches are proposed. One such possibility is the conversion of gravitons into photons in the presence of a magnetic field—a process known as the inverse Gertsenshtein effect~\cite{Gertsenshtein:1962kfm,Macedo:1983wcr,Cruise:2012zz,Dolgov:2012be,Ejlli:2018hke,Ejlli:2019bqj,Ejlli:2020fpt,Cembranos:2023ere,Palessandro:2023tee}. This mechanism is conceptually similar to axion-photon conversion in external magnetic fields, a well-established approach for probing axions, axion-like particles (ALPs) and dark photons~\cite{Raffelt:1987im,Roy:2023rjk,Poddar:2026wyd}.

Several prior studies have explored the feasibility of detecting high-frequency GWs via graviton-photon conversion in various environments, including cosmic magnetic fields~\cite{Kanno:2023fml,Dunsky:2025pvd}, the magnetic field near the galactic center~\cite{Ramazanov:2023nxz, Ito:2023nkq}, planetary magnetosphere~\cite{Liu:2023mll,Ito:2023nkq}, galactic neutron star populations~\cite{Dandoy:2024oqg,Ito:2023fcr}, blazar jets~\cite{Matsuo:2025blj},  electromagnetic cavities~\cite{Schenk:2025ria}, extragalactic magnetic fields~\cite{Domcke:2020yzq}, and those found in galaxy clusters~\cite{He:2023xoh}.

The giant elliptical galaxy Messier 87 (M87), located at the center of the Virgo cluster, hosts a supermassive black hole with mass $M_{\rm BH} \sim 6.5\times10^{9}M_\odot$ and exhibits one of the most powerful relativistic jets observed in the local Universe. Its exceptionally rich and structured magnetized environment, extending from the event-horizon scale to kiloparsec distances, makes M87 an ideal astrophysical laboratory for studying photon-boson conversion phenomena. The presence of large-scale ordered magnetic fields in the jet and accretion region provides favorable conditions for graviton–photon and axion–photon conversion processes. In particular, photon–axion mixing in magnetized plasmas has been explored in detail in~\cite{Nomura:2022zyy,Roy:2023rjk}. Similarly, photon–dark photon oscillations in structured magnetic backgrounds can lead to observable spectral distortions, as discussed in~\cite{Poddar:2026wyd}.  
Beyond its magnetized plasma environment, M87 is also expected to be embedded within a substantial dark matter halo, as indicated by observational and theoretical studies, and in some scenarios involving ultralight fields, wave-like effects on galactic scales can lead to the formation of solitonic core structures in such massive systems~\cite{Lacroix:2015lxa,HAWC:2023bti,Phoroutan-Mehr:2024cwd,Kar:2025ykb,Davies:2019wgi,Bar:2019pnz,Sarkar:2025tiy}.

In this paper, we explore the possibility of probing GWs over a wide frequency range, $10^{10} - 10^{27}~\mathrm{Hz}$, via graviton-photon conversion in the magnetic field of the M87 galaxy. 
Observations confirm the presence of magnetic fields in the central region of M87~\cite{EventHorizonTelescope:2021srq,Ro:2023fww,Kino:2015dha,Hada:2012fn,Hada:2015okc,VERITAS:2010udc,Kim:2018hul}. 
If gravitons from the cosmic background convert into photons through the inverse Gertsenshtein effect in the presence of magnetic fields, they may contribute to the observed electromagnetic spectrum. The spectral data, spanning from millimetre wavelengths to TeV gamma rays, have been analyzed and made publicly available by the Multi-Wavelength Working Group (MWL WG) in collaboration with the Event Horizon Telescope (EHT) as part of a broadband observational campaign of the M87 galaxy~\cite{EventHorizonTelescope-Multi-wavelengthscienceworkinggroup:2024xhy}. We adopt
the standard astrophysical model (as presented in the campaign paper) 
to interpret the spectrum, with the graviton-induced photon flux considered as an additional contribution.

By comparing the expected photon flux from graviton-photon conversions in a magnetic field with the observed electromagnetic background, we place upper limits on the characteristic strain amplitude of GWs. Our analysis yields improved constraints on the strain amplitude, $h_c$, achieving 1-5 order-of-magnitude 
improvement relative to existing astrophysical bounds.

The paper is organized as follows:  
In Sec.~II, we review the theoretical framework of graviton-photon conversion in magnetic fields, focusing on the inverse-Gertsenshtein effect.  
Sec.~III discusses the properties of the central region of the M87 galaxy, including its magnetic field and plasma profiles.  
In Sec.~IV, we discuss the amount of conversion of the gravitons to photons in M87 galaxy under the magnetic field and radiative plasma.
In Sec.~V, we estimate the theoretical photon flux resulting from graviton-photon conversion.  
Sec.~VI presents the observational emission spectrum and astrophysical modeling.  
Sec.~VII contains our derived constraints on the GW strain amplitude and the spectral density.  
We conclude with a summary and outlook in Sec.~VIII. 

\section{Formalism For Graviton to photon conversion in external magnetic fields }
In this section, we briefly review the inverse Gertsenshtein effect, a mechanism whereby gravitons can convert into photons in the presence of an external magnetic field.
Drawing upon the formalism established in Refs.~\cite{Dolgov:2012be,Ramazanov:2023nxz,Ejlli:2018hke}, we aim to estimate the probability for this graviton to photon ( $h \rightarrow \gamma$) transition over a given propagation distance. We consider the dynamics of gravitational and electromagnetic waves in flat spacetime.

Gravitons as well as gravitational waves are modeled as small perturbations ($h_{\mu\nu} \ll 1$) on top of a Minkowski background. Thus, the spacetime metric takes the form:
\begin{equation}
    g_{\mu\nu} = \eta_{\mu\nu} +\frac{2}{M_{Pl}} h_{\mu\nu},
\end{equation}
where the flat spacetime metric is defined as $\eta_{\mu\nu} = \mathrm{diag}(1,-1,-1,-1)$. Throughout, we work in natural units, setting $\hbar = c = 1$ and the reduced planck constant, $M_{pl}=2.44\times 10^{18}\rm GeV$ .

We begin with the quantum electrodynamics (QED) action describing photons minimally coupled to gravity, including the leading nonlinear Euler--Heisenberg correction,
\begin{equation}
\begin{aligned}
S_{\text{grav+EM}}
&= \int d^4x\,\sqrt{-g}\Bigg[
\frac{M_{\rm Pl}^2}{2}R
-\frac{1}{4}F_{\mu\nu}F^{\mu\nu}
+ A_\mu J^\mu \\
&\qquad
+\frac{\alpha^2}{90\,m_e^4}
\left(
\left(F_{\mu\nu}F^{\mu\nu}\right)^2
+\frac{7}{4}\left(F_{\mu\nu}\tilde F^{\mu\nu}\right)^2
\right)
\Bigg]
\end{aligned}
\end{equation}

where the electromagnetic field strength tensor and its dual are defined as
\begin{equation}
F_{\mu\nu} = \partial_\mu A_\nu - \partial_\nu A_\mu,
\qquad
\tilde F^{\mu\nu} = \frac{1}{2}\,\epsilon^{\mu\nu\alpha\beta}F_{\alpha\beta}.
\end{equation}


Varying the action with respect to \(A_\nu\) and keeping terms to linear order in \(h_{\mu\nu}\) and in the dynamical photon \(f_{\mu\nu}\) (treating any strong external field \(\bar F_{\mu\nu}\) as background) yields
\begin{equation}\label{Maxwell_lin}
\begin{aligned}
\partial_{\mu}\Big[F^{\mu\nu}-\frac{\alpha^2}{45m_e^4}\big(4F^2 F^{\mu\nu}+7(F\cdot\tilde F)\tilde F^{\mu\nu}\big)\Big]+J^{\nu}\\
=\partial_{\mu}\big(h^{\mu\beta}\eta^{\nu\alpha}\bar F_{\beta\alpha}-h^{\nu\beta}\eta^{\mu\alpha}\bar F_{\beta\alpha}\big),
\end{aligned}
\end{equation}
where \(F_{\mu\nu}=\bar F_{\mu\nu}+f_{\mu\nu}\), and on the right-hand side we have kept only the leading linear coupling between the metric perturbation \(h\) and the background field \(\bar F\). The electromagnetic current \(J^\nu\) includes the plasma response; in a cold plasma approximation one may write \(J^\nu=-\omega_{\rm pl}^2 A^\nu\) (in Coulomb gauge for the transverse components).
Variation of the gravitational part and linearization gives the usual linearized Einstein equation in Transverse-Traceless(TT) gauge,
\begin{equation}\label{Einstein_lin}
\Box h_{\mu\nu} = -\frac{2}{M_{\rm pl}}\,T_{\mu\nu}^{(em)} ,
\end{equation}
where \(T_{\mu\nu}^{(em)}\) is the electromagnetic energy-momentum tensor expanded to first order in dynamical photon fields \(f_{\mu\nu}\) and in the presence of the background \(\bar F_{\mu\nu}\). Keeping only the terms that are linear in the dynamical photon field \(f\) and linear in the background \(\bar F\) yields the source term responsible for graviton \(\leftrightarrow\) photon mixing,
\begin{equation}
\Box h_{ij} \simeq -\frac{\sqrt{2}}{M_{\rm pl}}\big(\bar B_i B_j + \bar B_j B_i\big),
\end{equation}
where \(i,j\) are spatial indices transverse to the propagation direction and \(\bar{\mathbf{B}}\) is the external (background) magnetic field. The numerical prefactor depends on the precise normalization of the metric perturbation; here we adopt the normalization consistent with \(g_{\mu\nu}=\eta_{\mu\nu}+2h_{\mu\nu}/M_{\rm pl}\).

We adopt the Coulomb gauge for photons, \(\nabla\cdot\mathbf{A}=0\), and the TT gauge for the gravitational perturbation. For a wave propagating along the \(z\)-axis we decompose the fields as
\begin{align}
A_i(\mathbf{x},t) &= \sum_{\lambda=\parallel,\perp} A_\lambda(z)\,\epsilon_i^\lambda\,e^{-i\omega t},\\
h_{ij}(\mathbf{x},t) &= \sum_{s=+,\times} h_s(z)\,e_{ij}^s\,e^{-i\omega t},
\end{align}
with \(\epsilon_i^\lambda\) the photon polarization vectors (orthonormal and transverse) and \(e_{ij}^s\) the graviton polarization tensors constructed from \(\epsilon_i^\lambda\) as
\(e^+_{ij}=\epsilon_i^\parallel\epsilon_j^\parallel-\epsilon_i^\perp\epsilon_j^\perp\),
\(e^\times_{ij}=\epsilon_i^\parallel\epsilon_j^\perp+\epsilon_i^\perp\epsilon_j^\parallel\).

Using \(\partial_t\to -i\omega\) and \(\Box\simeq (\omega+i\partial_z)(\omega-i\partial_z)\simeq 2\omega(\omega+i\partial_z)\) under the WKB approximation, the coupled linearized Eqs. \eqref{Maxwell_lin} and \eqref{Einstein_lin} for the slow \(z\)-dependent envelopes take the 
form \cite{Raffelt:1987im,Ejlli:2018hke,Lella:2024dus},

\begin{equation}\label{gr-photon-mode-eq}
\left(i\frac{d}{dz}+\omega\right)
\begin{pmatrix} h_+ \\ h_\times \\ A_\parallel \\ A_\perp \end{pmatrix}
=\mathcal{H}\begin{pmatrix} h_+ \\ h_\times \\ A_\parallel \\ A_\perp \end{pmatrix},
\end{equation}
where the mixing matrix $\mathcal{H}$ having the block form:
\begin{equation}
    \mathcal{H} =
    \begin{pmatrix}
        0 & \mathcal{H}_{g\gamma} \\
        \mathcal{H}_{g\gamma} & \mathcal{H}_{\gamma\gamma}
    \end{pmatrix}
\end{equation}

Here, both $\mathcal{H}_{g\gamma}$ and $\mathcal{H}_{\gamma\gamma}$ are $2 \times 2$ matrices. The matrix $\mathcal{H}_{\gamma\gamma}$ incorporates the dispersion of photons in a medium, while $\mathcal{H}_{g\gamma}$ captures the graviton-photon interaction induced by the presence of an external magnetic field.

The mixing component is given by:
\begin{equation}
    \mathcal{H}_{g\gamma} =
    \begin{pmatrix}
        \Delta_{g\gamma} \sin\phi & \Delta_{g\gamma} \cos\phi \\
        \Delta_{g\gamma} \cos\phi & -\Delta_{g\gamma} \sin\phi
    \end{pmatrix}
\end{equation}
where the angle $\phi$ defines the orientation of the external magnetic field with respect to the photon polarization vectors, such that $\cos\phi = \vec{B} \cdot \vec{\epsilon}_{\parallel} / B_T$, $\vec{B_T}$ is the transverse component of the magnetic field.

The strength of the mixing is quantified by:
\begin{equation}
    \Delta_{g\gamma} = \frac{B_T}{\sqrt{2} M_{\text{pl}}} 
\end{equation}

Photon propagation effects are encoded in the $\mathcal{H}_{\gamma\gamma}$ matrix:
\begin{equation}
    \mathcal{H}_{\gamma\gamma} =
    \begin{pmatrix}
        \Delta_{\gamma}^{\parallel} \cos^2\phi + \Delta_{\gamma}^{\perp} \sin^2\phi & (\Delta_{\gamma}^{\parallel} - \Delta_{\gamma}^{\perp}) \cos\phi \sin\phi \\
        (\Delta_{\gamma}^{\parallel} - \Delta_{\gamma}^{\perp}) \cos\phi \sin\phi & \Delta_{\gamma}^{\parallel} \sin^2\phi + \Delta_{\gamma}^{\perp} \cos^2\phi
    \end{pmatrix}
\end{equation}

Each of the terms $\Delta_{\gamma}^{\lambda}$, for $\lambda = \parallel, \perp$, includes contributions from plasma effects, QED corrections, and interactions with the cosmic microwave background (CMB):
\begin{equation}
    \Delta_{\lambda} = \Delta_{\text{pl}} + \Delta_{\text{QED}}^{\lambda} + \Delta_{\text{CMB}}
\end{equation}

The explicit forms of these contributions are:
\begin{align}
    \Delta_{\text{pl}} &= -\frac{\omega_{\text{pl}}^2}{2\omega}, \quad \omega_{\text{pl}}^2 = \frac{e^2 n_e(z)}{m_e}\label{plasma_term} \\
    \Delta_{\text{QED}}^{\lambda} &= k_{\lambda} \frac{4 \alpha^2 B_T^2(z) \omega}{45 m_e^4}\label{QED_term} \\
    \Delta_{\text{CMB}} &= \frac{44\pi^2 \alpha^2 T_{\text{CMB}}^4 \omega}{2025 m_e^4}\label{cmb_term}
\end{align}

Here, $k_{\lambda} = 7/2$ for $\lambda = \parallel$ and $k_{\lambda} = 2$ for $\lambda = \perp$. The plasma frequency $\omega_{\text{pl}}$ depends on the electron number density $n_e$, which, for instance, in a galactic environment such as the Milky Way, where typically $n_e \sim 10^{-2}\,\text{cm}^{-3}$, yields $\omega_{\text{pl}} \sim \mathcal{O}(10^{3})\,\text{Hz}$. The CMB temperature is approximately $T_{\text{CMB}} \simeq 2.73\,\text{K}$~\cite{Fixsen:2009ug}. It should be noted that the evolution Eq.~\eqref{gr-photon-mode-eq} remains valid provided the mixing species are relativistic, i.e., when the dispersion relation satisfies $k \simeq \omega$. This condition holds for all analyses in this work. Here, we mainly focus on frequencies 
$f \in [10^{10},\,10^{27}]\,\mathrm{Hz}$, corresponding to photon energies 
$40\,\mu\mathrm{eV} \lesssim \omega \lesssim 40\,\mathrm{GeV}$. 
Under the typical plasma conditions of the M87 environment, the plasma frequency is such that the relativistic approximation $k \simeq \omega$ remains valid throughout the considered parameter range. Furthermore, Eq.~\eqref{gr-photon-mode-eq} remains applicable for photon energies up to $\omega \lesssim 100\,\mathrm{TeV}$, since attenuation effects arising from scattering with CMB photons and the extra-galactic background light (EBL) are negligible in this domain (see Refs.~\cite{Mirizzi:2009aj,Dobrynina:2014qba,Kartavtsev:2016doq} for related discussions).

We now seek solutions of Eq.~\eqref{gr-photon-mode-eq}. 
The mode equation~\eqref{gr-photon-mode-eq} does not admit a closed-form analytical solution in the presence of a spatially varying magnetized medium. Consequently, we adopt a numerical approach, imposing initial conditions corresponding to unpolarized gravitons in a region where their distribution is assumed to be homogeneous. 
For completeness, we also provide the analytical solution of the mode equation under the simplifying assumption of a constant and homogeneous magnetic field in Appendix~\ref{appenA}.

Starting from Eq.~\eqref{gr-photon-mode-eq}, we introduce a state vector, construct the mixing Hamiltonian, propagate the state, 
and extract the photon conversion probabilities in the observer's polarization basis.
We define the four-component state vector
\begin{equation}
\Psi(z) \equiv \bigl( h_{+}(z),\, h_{\times}(z),\, A_{\parallel}(z),\, A_{\perp}(z) \bigr)^{T},
\end{equation}
where $A_{\parallel}$ and $A_{\perp}$ denote photon polarization components defined with respect to the local transverse magnetic-field direction, and $h_{+}$ and $h_{\times}$ are the two linear graviton polarizations.

The propagation equation can be written in Schr\"odinger-like form
\begin{equation}\label{schrodinger_like}
i\frac{d}{dz}\Psi(z) = \mathcal{H}(z)\Psi(z) - \omega\,\Psi(z),
\end{equation}
or equivalently,
\begin{equation}\label{mode_evolution}
i\frac{d}{dz}\Psi(z) = \mathcal{H}_{\rm eff}(z)\Psi(z),
\qquad
\mathcal{H}_{\rm eff}(z) \equiv \mathcal{H}(z) - \omega\,\mathbb{I}_{4}.
\end{equation}

The Hamiltonian $\mathcal{H}(z)$ depends explicitly on the propagation coordinate through the spatial profiles of the magnetic field strength $B(z)$, the electron number density $n_e(z)$, and the magnetic-field orientation angle $\phi(z)$. The effective Hamiltonian is Hermitian at each position, ensuring unitary evolution.

The formal solution of Eq.~\eqref{mode_evolution} is
\begin{equation}
\begin{split}
\Psi(z_f) &= \mathcal{U}(z_f,z_i)\Psi(z_i), \\
\text{where},\quad \mathcal{U}(z_f,z_i) &=
\mathcal{T}\exp\!\Bigg[
-i\!\int_{z_i}^{z_f}
\mathcal{H}_{\rm eff}(z)\,dz
\Bigg].
\end{split}
\end{equation}
where $\mathcal{T}$ denotes path ordering.

Numerically, the propagation path is discretized into $N$ uniform steps of size $\Delta z$. Over each step the evolution operator is constructed by diagonalizing the local effective Hamiltonian and exponentiating its eigenvalues, yielding the stepwise evolution
\begin{equation}
\Psi(z+\Delta z) \simeq
\exp\!\left[-i\,\mathcal{H}_{\rm eff}(z)\,\Delta z\right]\Psi(z).
\end{equation}
The full evolution is obtained as an ordered product of these stepwise operators.

We propagate independently the two pure graviton initial states
\begin{equation}
\Psi^{(+)}(z_i) =
\begin{pmatrix}
1\\0\\0\\0
\end{pmatrix},
\qquad
\Psi^{(\times)}(z_i) =
\begin{pmatrix}
0\\1\\0\\0
\end{pmatrix},
\end{equation}
obtaining the final states $\Psi^{(+)}(z_f)$ and $\Psi^{(\times)}(z_f)$.

From the evolved state
$\Psi = (h_+,h_\times,A_\parallel,A_\perp)^{T}$,
the photon amplitudes in the interaction basis are extracted as
\begin{equation}
A_\parallel = \Psi_3,
\qquad
A_\perp = \Psi_4.
\end{equation}

The photon conversion probabilities for each initial graviton polarization are
\begin{align}
P_{h_+ \to A_\parallel} &= |A_\parallel^{(+)}|^2, &
P_{h_+ \to A_\perp} &= |A_\perp^{(+)}|^2,\\[2mm]
P_{h_\times \to A_\parallel} &= |A_\parallel^{(\times)}|^2, &
P_{h_\times \to A_\perp} &= |A_\perp^{(\times)}|^2.
\end{align}
Since the incoming gravitational-wave background is assumed to be unpolarized, the
physically relevant conversion probabilities are obtained by averaging over the two initial
graviton polarizations,
\begin{align}
P^{\rm unpolarized}_{h \to A_\parallel}
&=
\frac{1}{2}\left(
P_{h_+\to A_\parallel} + P_{h_\times\to A_\parallel}
\right),\\[2mm]
P^{\rm unpolarized}_{h \to A_\perp}
&=
\frac{1}{2}\left(
P_{h_+\to A_\perp} + P_{h_\times\to A_\perp}
\right).
\end{align}
Finally, since no polarization information of the converted photons is assumed to be
experimentally accessible, the total graviton–photon conversion probability is obtained by
summing over the two orthogonal photon polarization states,
\begin{equation}\label{Eq_total_prob}
P^{\rm total}_{h \to \gamma}
=
P^{\rm unpolarized}_{h \to A_\parallel}
+
P^{\rm unpolarized}_{h \to A_\perp}.
\end{equation}
This quantity determines the total number of photons produced via graviton–photon conversion and constitutes the relevant input for the calculation of the observable photon flux
at Earth.
\begin{figure*}
\centering
    \includegraphics[width=8.5cm, height =6.5cm]{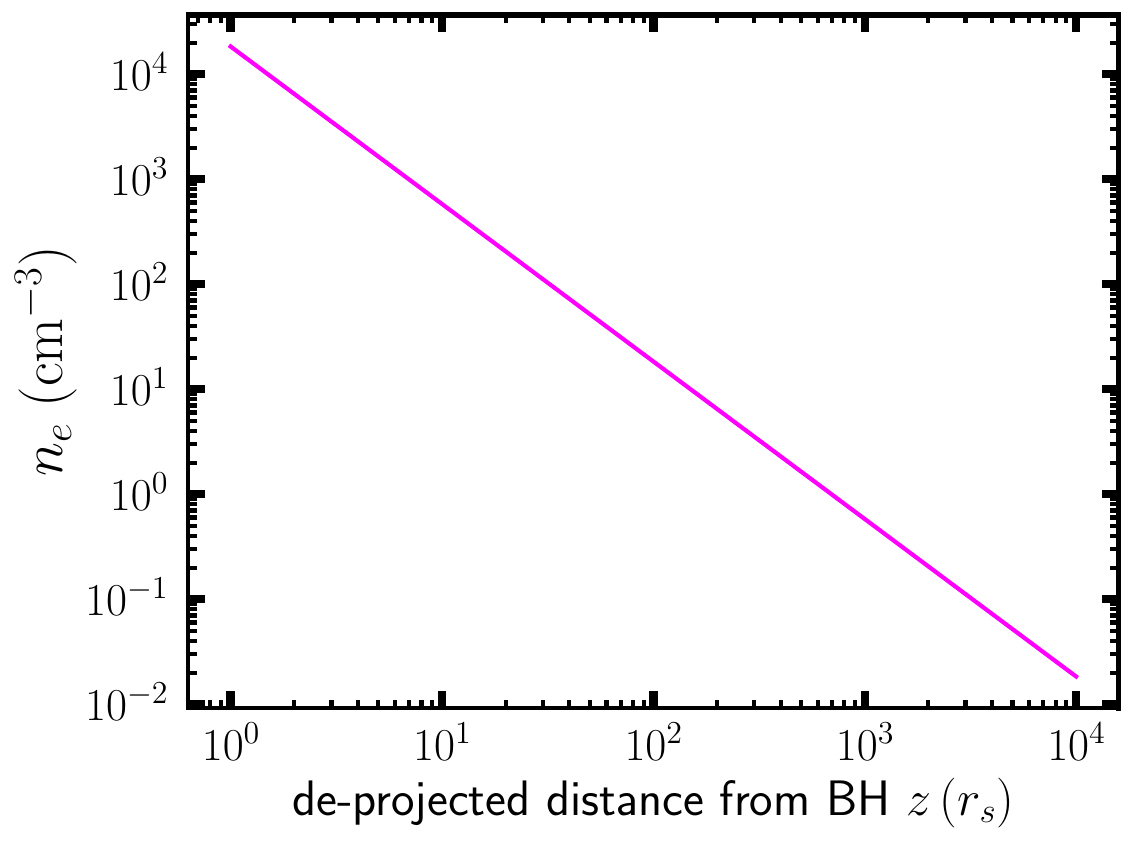}
    \includegraphics[width=8.5cm, height =6.5cm]{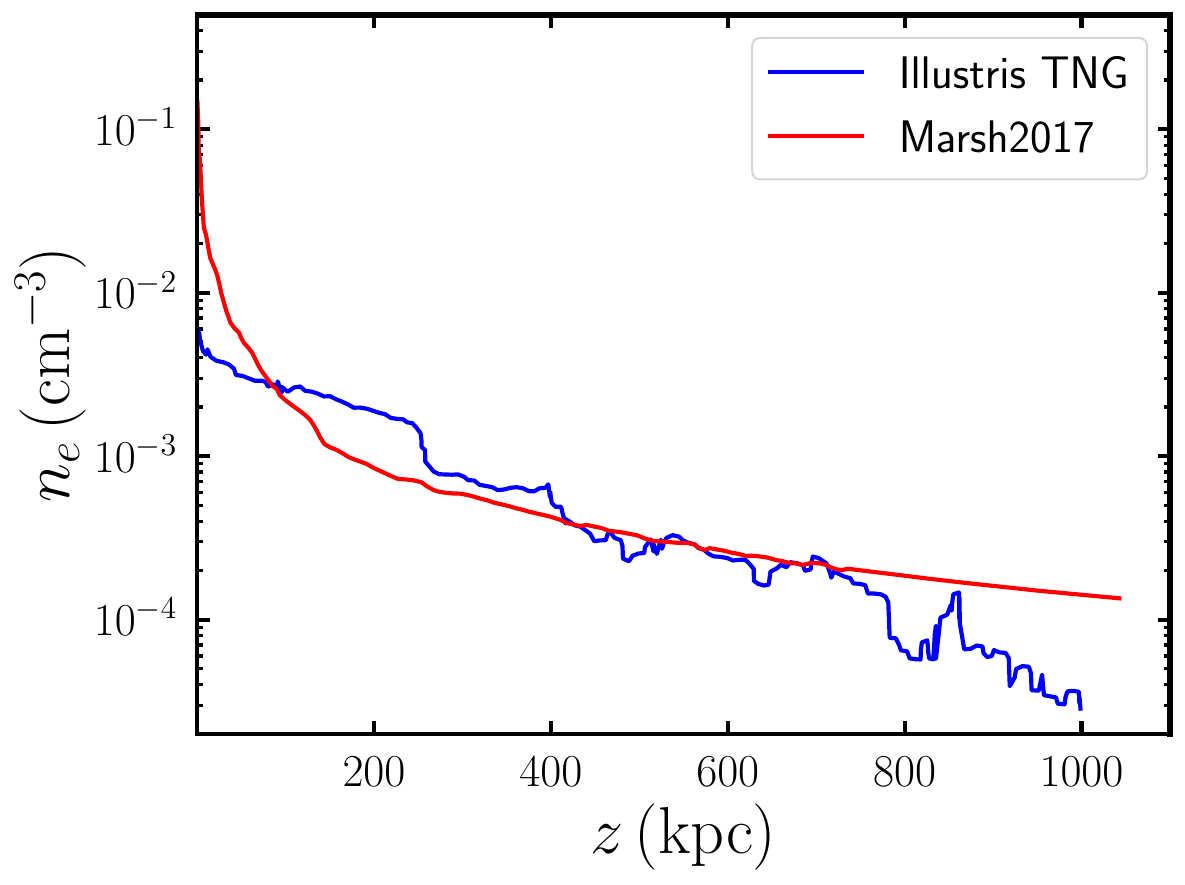}
    \caption{Profiles of the free-electron number density in the M87 galaxy. The left panel shows the plasma density in the near-horizon region of the supermassive black hole according to Eq.~\eqref{plasma_prfl}. The right panel displays the plasma density profile on larger scales: the blue curve corresponds to results from the IllustrisTNG300 simulations adopted from Ref.~\cite{Ning:2024eky}, while the red curve represents the profile inferred from  Ref.~\cite{Marsh:2017yvc}.}
    \label{fig:electron_density}    
\end{figure*}
\section{Electron density and Magnetic Field profiles of M87}
\begin{figure*}
\centering
    \includegraphics[width=8.5cm, height =6.5cm]{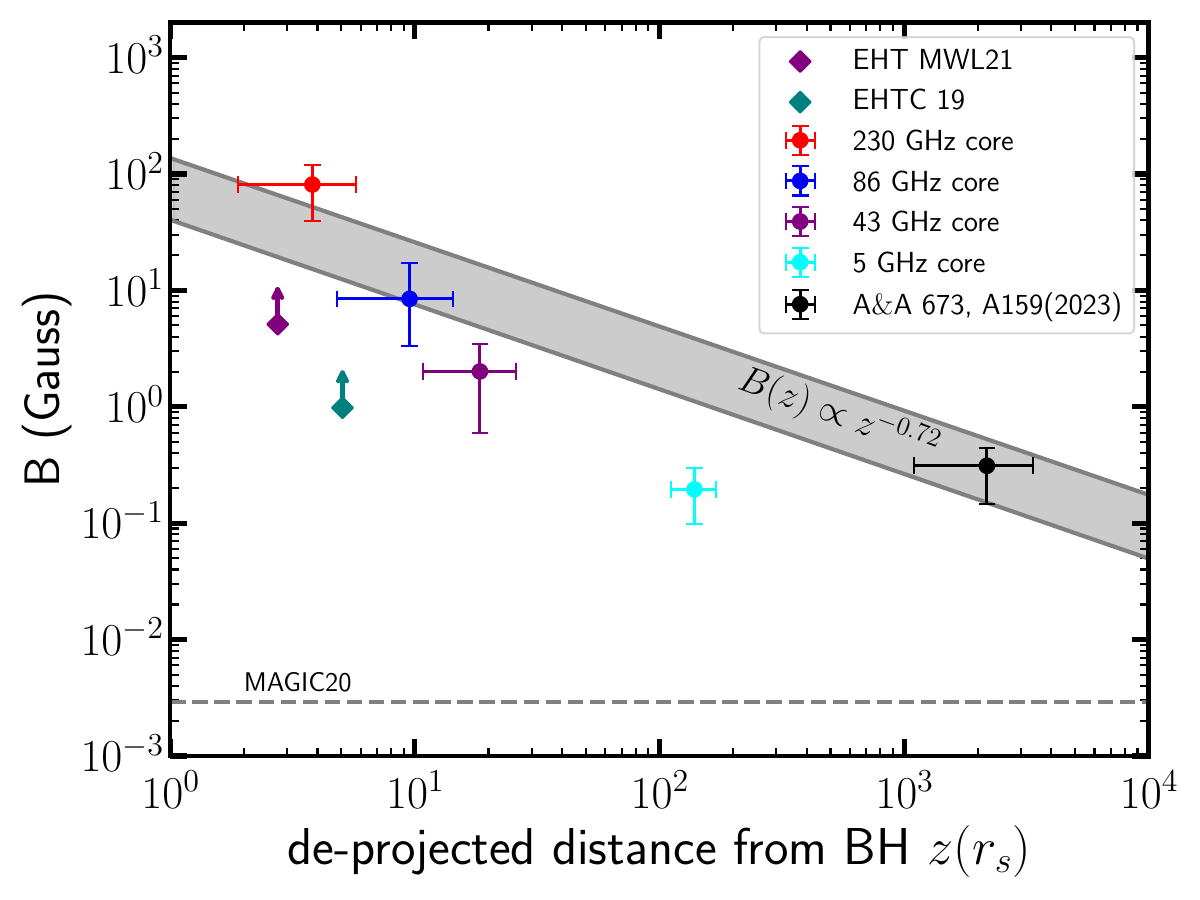}
    \includegraphics[width=8.7cm, height =6.6cm]{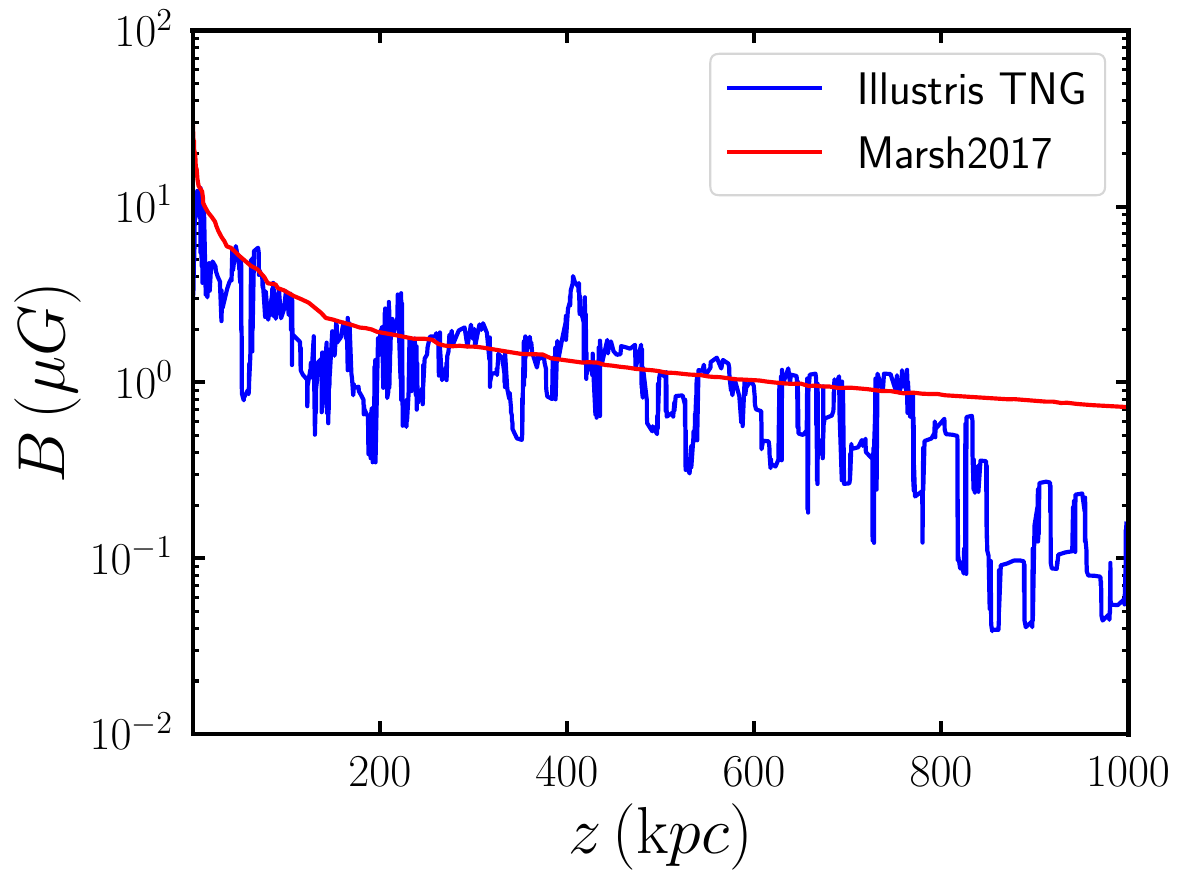}
    \caption{Radial profiles of the magnetic field strength in the M87 galaxy. The left panel shows near-horizon magnetic-field estimates inferred from VLBI, KVN, and multi-wavelength observations, together with the parametrized profile defined in Eq.~\eqref{magnetic_prfl} and is adopted from~\cite{Ro:2023fww}. The right panel presents large-scale magnetic-field profiles, with the blue curve corresponding to Illustris TNG300 simulations \cite{Ning:2024eky} and the red curve representing the profile inferred from rotation-measure observations of M87 \cite{Marsh:2017yvc}.}
    \label{fig:mag_field}    
\end{figure*}
The graviton-photon conversion probability depends upon the magnetic field and free electron density of the medium. These two quantities play a critical role in our analyses.
In this section, we construct realistic profiles for these quantities in the M87 system. The galaxy M87 is a massive elliptical galaxy located in the Virgo Cluster and hosts a SMBH at its center. The presence of strong magnetic fields and a hot plasma in the vicinity of the central black hole, M87*, is well established through a variety of theoretical and observational models describing the inner region of M87~\cite{EventHorizonTelescope:2019uob,EventHorizonTelescope:2019jan,EventHorizonTelescope:2019ths,EventHorizonTelescope:2019pgp,EventHorizonTelescope:2019ggy,EventHorizonTelescope:2021srq}. The plasma environment surrounding the M87* is frequently modeled using simplified accretion prescriptions. A widely used representation is the spherical accretion model, which assumes a stationary and radially symmetric inflow of material. For a constant mass accretion rate, expressed as 
\(\dot{M} = 4\pi r^2 \rho v_r\), 
and adopting a free-fall velocity scaling \(v_r \propto r^{-1/2}\), the resulting mass density follows \(\rho \propto r^{-3/2}\). Consequently, the electron number density can be approximated by a power-law distribution~\cite{Rybicki1979RadiativePI,Quataert:2002xn,Yuan:2014gma,EventHorizonTelescope:2019pgp,Roy:2023rjk,Hada:2024icg}:  
\begin{equation}\label{plasma_prfl}
n_e = n_{0}\left(\frac{z}{r_{ph}}\right)^{-3/2},
\end{equation}
where $n_0$ denotes the electron number density near the photon sphere of M87* and \(r_{ph}\) represents the photon sphere radius, \(r_{ph} = \tfrac{3}{2}\,r_s\), with the Schwarzschild radius given by $r_s = 2 G M_{\rm BH}/c^{2}$. The mass of the black hole is estimated to be $M_{BH} \sim 6.5 \times 10^{9} M_{\odot}$.

The plasma in the immediate vicinity of the SMBH's event horizon is hot and magnetized, with electron temperatures on the order of \(T_e \sim 10^9\,\mathrm{K}\) and magnetic field strengths of about \(1\text{--}30\,\mathrm{G}\)~\cite{EventHorizonTelescope:2021srq}. For M87*, millimeter-wave observations imply electron densities of \(n_e \sim 10^4\text{--}10^7\,\mathrm{cm^{-3}}\) at radial distances of roughly \((5\text{--}10)\,r_g\) (with \(r_g = GM_{BH}/c^2\) being the gravitational radius). These parameters correspond to sub-Eddington accretion rates, \(\dot{M} \sim 10^{-5}\,M_\odot\,\mathrm{yr^{-1}}\), indicating a hot, optically thin, and magnetically dominated plasma surrounding the black hole~\cite{EventHorizonTelescope:2019pgp,Hada:2024icg}. To complement the analytic plasma density model described above, we also incorporate cosmologically motivated electron density distributions from Ref.~\cite{Ning:2024eky}, derived from the \textit{IllustrisTNG300} magnetohydrodynamic simulations~\cite{Marinacci:2017wew,Pillepich:2019bmb,Nelson:2019jkf}. The Illustris project and its successor, IllustrisTNG, are designed to simulate the evolution of the Universe from shortly after the Big Bang to the present day, self-consistently tracking the interplay between dark matter, baryonic matter, and supermassive black holes within the standard cosmological paradigm. The TNG300 simulation, with a comoving box size of 300 Mpc, offers an exceptional statistical sample for investigating massive galaxies and galaxy clusters, including systems analogous to M87. In Virgo-like cluster counterparts with total masses of $(6.3 \pm 0.9)\times10^{14},\mathrm{M}_\odot$, the electron number density reaches values of order $10^{-2},\mathrm{cm}^{-3}$ within the inner tens of kiloparsecs and decreases to approximately $10^{-3},\mathrm{cm}^{-3}$ at radial distances of several hundred kiloparsecs. The analytical electron density profile given in Eq.~\eqref{plasma_prfl} is shown in the left panel of Fig.~\ref{fig:electron_density}, while the large-scale electron density profile, reproduced from Ref.~\cite{Ning:2024eky} in the case of Illustris TNG (blue colour) and Ref.~\cite{Marsh:2017yvc} in the case of Marsh et.al. (red colour), is displayed in the right panel of the same figure. In the present analysis, we combine these two descriptions: the inner region, extending up to $10^4{r_s}$ from the supermassive black hole, is modeled using the power-law profile of Eq.~\eqref{plasma_prfl}, and the outer region, which is described by the simulated large-scale profile.

The structure and magnitude of the magnetic field in the emission from M87 remain subject to considerable uncertainty, as different observational techniques probe distinct spatial scales and rely on varying physical assumptions. In this work, we adopt a distance-dependent magnetic-field profile along the SMBH following Ref.~\cite{Hada:2024icg}. Here, the magnetic field is inferred from high-resolution very long baseline interferometry (VLBI) observations,
carried out using the Korean VLBI Network (KVN), the VERA Array (KaVA), and the Very Long Baseline Array (VLBA). 
 
In the left panel of Fig.~\ref{fig:mag_field}, we present representative estimates and constraints on the magnetic-field strength spanning from the immediate vicinity of the event horizon to the outer regions of the M87 galaxy.

VLBI core analyses performed at multiple frequencies (shown as colored circles and diamonds in left panel of Fig.~\ref{fig:mag_field}) infer magnetic-field strengths in the range $B \sim 0.1$--$10\,\mathrm{G}$ near the radio core, based on synchrotron self-absorption and core-shift measurements \cite{Kino:2015dha,Hada:2012fn,Hada:2015okc,VERITAS:2010udc,Kim:2018hul,Zamaninasab:2014dhz,Jiang:2021inh}. On larger spatial scales, spectral energy distribution (SED) modeling of multi-wavelength emission provides complementary constraints, typically indicating weaker fields of $B \lesssim 10^{-2}\,\mathrm{G}$, as suggested by MAGIC observations (dashed gray line) \cite{MAGIC:2020gbb}.

Additional information on the magnetized plasma in the immediate vicinity of the black hole is provided by polarimetric observations from the Event Horizon Telescope (EHT), which probe horizon-scale regions and indicate magnetic-field strengths of order $B \sim (1–10)\mathrm{G}$ \cite{EventHorizonTelescope:2021dvx}. Complementarily, Ref.~\cite{Ro:2023fww} infers magnetic fields in the range $B \sim 0.1$–$1,\mathrm{G}$ by studying the spatial variation of the synchrotron spectral index across the extended emission environment of the M87 galaxy, suggesting a power-law behavior $B(z) \propto z^{-0.72}$. The shaded region in the left panel of Fig.~\ref{fig:mag_field} illustrates the extrapolation of this magnetic-field profile. The widespread nature of these magnetic-field estimates underscores the substantial systematic uncertainties associated with modeling the magnetized emission environment of M87. 

The magnetic-field strength can be modeled as a function of the distance $z$ from the central black hole according to \cite{Ro:2023fww,Hussein:2025llu},
\begin{equation}\label{magnetic_prfl}
B_T(z)
= (0.3\text{--}1)\,\mathrm{G}
\left(\frac{z}{900\,r_s}\right)^{-0.72}
\,\,\,
\end{equation}
where $r_s$ denotes the Schwarzschild radius of M87*.
In the right panel of Fig.~\ref{fig:mag_field}, we present magnetic-field profiles inferred from numerical simulation studies~\cite{Marsh:2017yvc,Ning:2024eky}. These simulations indicate that the magnetic field strength reaches values of order \(\sim10~\mu\mathrm{G}\) in the central few kiloparsec region and remains at an average level of \(\sim1~\mu\mathrm{G}\) out to radial distances of several hundred kiloparsecs. Such an extended magnetized environment provides favorable conditions for graviton–photon conversion processes due to the large coherence length. 

\begin{figure*}
    \centering
    \includegraphics[width=0.7\linewidth]{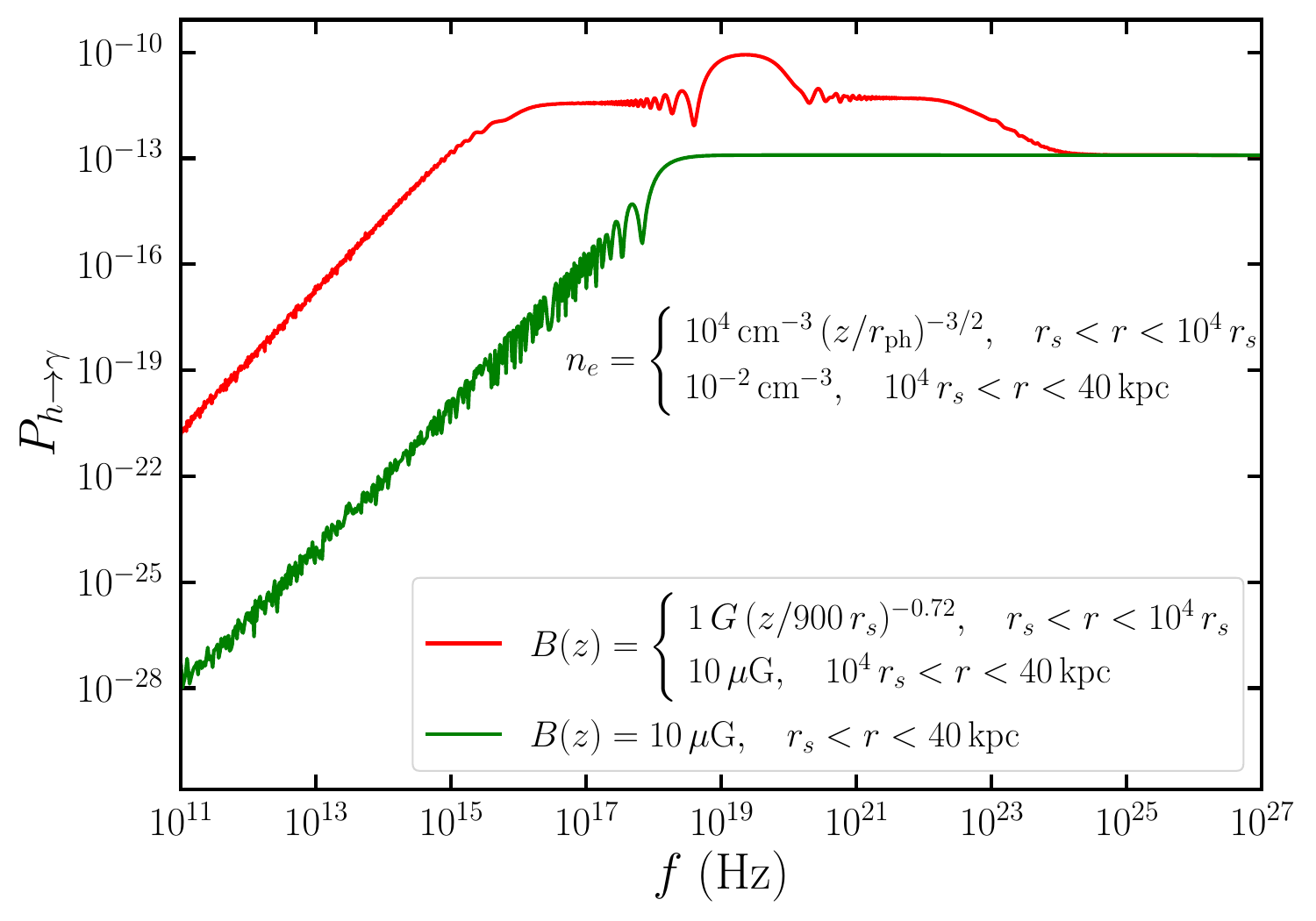}
    \caption{Total unpolarized graviton–to–photon conversion probability as a function of photon frequency. The red curve shows the numerically computed probability including the distance-dependent plasma and magnetic-field profiles, while the green curve corresponds to the case of constant magnetic field and plasma density, reproducing the analytical result derived in Appendix~\ref{appenA}.}
    \label{fig:num_prob}
\end{figure*}
\section{The graviton-photon conversion probability in M87 }

In this section, we investigate graviton--photon oscillations in the magnetized environment of the M87 galaxy. As shown in Fig.~\ref{fig:mag_field}, the magnetic-field strength is largest in the near-horizon region of the central supermassive black hole, M87*. We consider the oscillation region to extend out to a distance of \(z = 40\,\mathrm{kpc}\), beyond which the magnetic field decreases rapidly and the oscillation probability becomes negligible.

The plasma contribution enters through Eq.~\eqref{plasma_term}, for which we adopt the electron density profile shown in Fig.~\ref{fig:electron_density}. In addition, the QED vacuum polarization and the cosmic microwave background (CMB) terms, given in Eqs.~\eqref{QED_term} and \eqref{cmb_term}, respectively, can be explicitly evaluated in the M87 environment.

For the electron number density, we follow the profile displayed in the left panel of Fig.~\ref{fig:electron_density} in the vicinity of the black hole, corresponding to the region \(r_s < z < 10^{4} r_s\), and assume an average value of \(n_e \simeq 10^{-2}\,\mathrm{cm}^{-3}\) for larger distances, \(10^{4} r_s < z < 40\,\mathrm{kpc}\). With these assumptions, the plasma contribution to the mixing matrix takes the form
\begin{equation}\label{delta_plasma}
\begin{aligned}
\Delta_{\rm pl}
\simeq\;&
-1.1\,
\left(\frac{\omega}{\mathrm{GeV}}\right)^{-1}
\left(\frac{n_e(z)}{10^{4}\,\mathrm{cm}^{-3}}\right)
\,\mathrm{kpc}^{-1},\text{where}\quad \\[3pt]
&\frac{n_e(z)}{\mathrm{cm}^{-3}}=
\begin{cases}
10^{4}\,
\left(\dfrac{z}{r_{\rm ph}}\right)^{-3/2},
& r_s < z < 10^{4}\,r_s, \\[3pt]
10^{-2}\,,
& 10^{4}\,r_s < z < 40\,\mathrm{kpc}.
\end{cases}
\end{aligned}
\end{equation}

We adopt a similar piecewise prescription for the magnetic-field profile, as illustrated in both panels of Fig.~\ref{fig:mag_field}. In the region close to M87*, we use the parametrized magnetic-field profile given in Eq.~\eqref{magnetic_prfl}, which is also shown as the shaded region in the left panel of Fig.~\ref{fig:mag_field}. At larger distances, we assume an average magnetic-field strength of \(B \simeq 10\,\mu\mathrm{G}\), motivated by simulation-based profiles at large radii (shown by the red and blue curves in the right panel of Fig.~\ref{fig:mag_field}).

The dominant magnetic-field dependence enters through the graviton--photon mixing term,
\begin{equation}\label{delta_mixing}
\begin{aligned}
\Delta_{g\gamma}
\simeq\;&
8.8\times10^{-9}
\left(\frac{B_T(z)}{10\,\mu\mathrm{G}}\right)
\,\mathrm{kpc}^{-1},\,\,\text{where} \\[4pt]
&
B_T(z) =
\begin{cases}
1\,\mathrm{G}
\left(\dfrac{z}{900\,r_s}\right)^{-0.72},
& r_s < z < 10^{4}\,r_s, \\[4pt]
10\,\mu\mathrm{G}\,\,\,,
& 10^{4}\,r_s < z < 40\,\mathrm{kpc}.
\end{cases}
\end{aligned}
\end{equation}

The same magnetic-field profile also contributes to the QED correction,
\begin{equation}\label{delta_qed}
\Delta_{\rm QED}^{\lambda}
\simeq
4.5 \times 10^{-9}\,
k_{\lambda}\,10^3\,
\left(\frac{\omega}{\mathrm{GeV}}\right)
\left(\frac{B_T(z)}{1\,\mu\mathrm{G}}\right)^2
\,\mathrm{kpc}^{-1},
\end{equation}
while the contribution from the cosmic microwave background is given by
\begin{equation}\label{delta_cmb}
\Delta_{\rm CMB}
\simeq
8.7 \times 10^{-8}
\left(\frac{\omega}{1\,\mathrm{GeV}}\right)
\,\mathrm{kpc}^{-1}.
\end{equation}
Here, \(k_{\lambda} = 7/2\) for photons polarized parallel to the magnetic field and \(k_{\lambda} = 2\) for the perpendicular polarization.

Since we consider photon frequencies in the range \(10^{10}\text{--}10^{27}\,\mathrm{Hz}\), both the QED and CMB contributions become significant and must be consistently included in the analysis.

We now present the numerical results for the graviton--photon conversion probability obtained by solving the coupled propagation Eqs.. In our analysis, the initially unpolarized graviton modes, \(h_{+}\) and \(h_{\times}\), can convert into photons with polarization states parallel and perpendicular to the external magnetic field. We compute the total conversion probability by summing over the final photon polarizations and averaging over the initial graviton polarizations.

In Fig.~\ref{fig:num_prob}, the total unpolarized graviton--to--photon conversion probability (via Eq.~\eqref{Eq_total_prob}) is shown by the red curve. This result is obtained by including both the plasma contribution and the magnetic-field effects, as encoded in Eqs.~\eqref{delta_plasma}, \eqref{delta_mixing}, \eqref{delta_qed}, and \eqref{delta_cmb}, and by adopting the piecewise distance-dependent profiles for the plasma density and magnetic field over the two characteristic distance scales discussed in the previous section. 

For comparison, we also present the transition probability obtained by assuming a constant value for the magnetic field ($B=10 \mu G$) and a spatially constant plasma density with $n_e=10^{-2} \rm cm^{-3}$. This case is depicted by the green curve in Fig.~\ref{fig:num_prob}. As expected, the green curve closely follows the analytical result derived in Appendix~\ref{appenA}, where both the magnetic field and the plasma density are taken to be constant. The numerical result therefore provides a nontrivial consistency check by reproducing the analytical probability shown in the left panel of Fig.~\ref{fig:analytical_prob_osclength}. 
It is worthwhile to compare this probability with that reported in Ref.~\cite{Lella:2024dus} where conversion takes place inside Milky Way magnetic field. For the benchmark parameters considered above, the graviton-to-photon conversion probability in M87 is approximately $1.2\times10^{-13}$ for frequencies $f \gtrsim 10^{18},{\rm Hz}$ (see Eq.~\ref{prob_analytical} and the green curve in Fig.~\ref{fig:num_prob}), which is about $1600$ times larger than the corresponding value reported in Eq.~(19) of Ref.~\cite{Lella:2024dus}. This enhancement arises primarily from the larger propagation distance, $z$, considered in our analysis, together with the stronger average magnetic field strength in M87 compared to that of the Milky Way. Due to this, M87 is expected to provide stronger bounds on the gravitational-wave energy density compared to the Milky Way galaxy.

Furthermore, accounting for the spatial variation of the magnetic field and plasma density increases the conversion probability by an additional four to six orders of magnitude relative to the constant-field approximation, as shown by the solid red curve in Fig.~\ref{fig:num_prob}. The resulting probability is therefore significantly larger than that obtained for the Milky Way over the entire frequency range considered, largely owing to the presence of strong magnetic fields in the inner regions of M87. Consequently, M87 is expected to yield substantially stronger constraints on the gravitational-wave energy density and the characteristic strain than those derived from the Milky Way.

A distinctive feature of the red curve 
around $f\sim 10^{19}-10^{20}$ Hz is the appearance of localized structures and changes in slope at specific distances and photon frequencies. These features arise from the strong magnetic field in the inner region of the M87 environment and from the transition between different radial regimes in the magnetic-field and plasma-density profiles.
Overall, we find that the graviton--photon conversion probability increases with increasing photon frequency in both cases. However, the inclusion of realistic, spatially varying plasma and magnetic-field profiles leads to richer phenomenology, highlighting the importance of environmental effects in accurately modeling graviton--photon oscillations in astrophysical settings.

\section{Photon flux from gravitons}
A stochastic background of gravitons or GWs can be described statistically in terms of its energy density spectrum. The quantity most commonly used to characterize such a background is the dimensionless spectral energy density parameter~\cite{Maggiore:2000gv}
\begin{equation}
    \Omega_{\rm gw}(f) = \frac{1}{\rho_c} \frac{d\rho_{\rm gw}}{d\log f}~,
\end{equation}
where $\rho_{\rm gw}$ denotes the energy density of gravitational waves 
, and $\rho_c$ is the critical energy density of the Universe, defined as
\begin{equation}
    \rho_c = \frac{3 H_0^2}{8\pi G_N}~.
\end{equation}
Here, $H_0$ is the current Hubble expansion rate, conventionally written as $H_0 = h \times 100~\rm km\,s^{-1}\,Mpc^{-1}$, where $h$ encodes the observational uncertainty($h\sim 0.67-0.73$). $G_N$ is Newton's gravitational constant, and $f$ denotes the gravitational wave frequency. Since the uncertainty in $H_0$ is unrelated to the intrinsic properties of the GW background, theoretical predictions are often quoted in terms of $h^2\Omega_{\rm gw}(f)$, which is independent of the precise value of the Hubble parameter. 

While $\Omega_{\rm gw}(f)$ provides a convenient measure of the energy content of a stochastic GW background, its interaction with detectors and astrophysical environments is more naturally described in terms of metric perturbations. In this context, it is useful to introduce the characteristic strain amplitude $h_c(f)$, which quantifies the typical amplitude of GW fluctuations per logarithmic frequency interval. The ensemble-averaged metric perturbations satisfy the relation~\cite{Maggiore:2000gv}
\begin{equation}
    \langle h_{ij}(t)\,h^{ij}(t) \rangle
    =2
    \int_0^\infty d(\log f)\, h_c^2(f),
\end{equation}
where the angular brackets denote an average over realizations of the stochastic background. The characteristic strain is related to the  power spectral density $S_h(f)$ via
\begin{equation}
    h_c^2(f)
    =
    2f\,S_h(f),
\end{equation}
while the energy density spectrum can be written in terms of $S_h(f)$ as
\begin{equation}
    \Omega_{\rm gw}(f)
    =
    \frac{4\pi^2}{3H_0^2}\,
    f^3 S_h(f).
\end{equation}
Combining these relations yields a direct connection between the GW energy density and the strain amplitude,
\begin{equation}
    \Omega_{\rm gw}(f)
    =
    \frac{2\pi^2}{3H_0^2}\,
    f^2 h_c^2(f),
\end{equation}
which allows one to translate between cosmological descriptions of the GW background and observationally relevant strain spectra.

The energy density carried by the background gravitons per logarithmic frequency interval is therefore given by
\begin{equation}
    \frac{d\rho_{\rm gw}}{d\log f}
    =
    \rho_c\,\Omega_{\rm gw}(f)
    =
    2\pi^2 M_{\rm Pl}^2\,
    f^2 h_c^2(f),
\end{equation}
where $M_{\rm Pl}$ denotes the reduced Planck mass.
This quantity represents the incoming graviton flux incident on a localized astrophysical environment and constitutes the relevant source term for graviton--photon conversion processes.

In regions permeated by strong magnetic fields and plasma, gravitons can convert into photons through graviton--photon mixing. Such conditions are naturally realized 
in the magnetized atmospheres surrounding SMBH. In these environments, the magnetic field strength and plasma density vary with radius, leading to a position-dependent conversion probability. Denoting by $P^{\rm total}_{h\to\gamma}$ the total graviton--photon conversion probability accumulated along the graviton trajectory, the electromagnetic radiation produced by this mechanism can be quantified in terms of a photon flux.

Accounting for geometric dilution between the conversion region and the observer, the photon flux at Earth is given by~\cite{Ito:2023fcr,Matsuo:2025blj,Romano:2016dpx}~\footnote{The reduction by one power of $f$ arises because the particle number flux is obtained by dividing the energy density flux by the energy per quantum.} 
\begin{equation}\label{grav_induced_phflux}
    \Phi_{h \to \gamma}(f)
    =
    2\pi^2\,
    \frac{R^2}{d^2}\,
    M_{\rm Pl}^2f h_c^2(f)\,
    P^{\rm total}_{h\to\gamma},
\end{equation}
where $R$ characterizes the spatial extent of the region in which graviton--photon conversion is efficient, and $d$ denotes the distance between the conversion region (i.e. in our case, the M87 galaxy) and the observer. Throughout our analysis, we assume a distance of $d=16.8\,{\rm Mpc}$~\cite{EventHorizonTelescope:2019ths}. The parameter $R$ effectively sets the oscillation length contributing to the accumulated conversion probability. In the case of the M87 galaxy, we adopt a fiducial value $R \simeq 40~\mathrm{kpc}$, corresponding to the extent of the magnetized environment surrounding the central supermassive black hole. This choice is motivated by observational and theoretical studies indicating that the magnetic field strength in M87 decreases rapidly with distance from the central region, transitioning from the strongly magnetized jet and inner galactic environment to the more weakly magnetized intergalactic medium. Beyond scales of order tens of kiloparsecs, the magnetic field strength is expected to fall sufficiently that graviton--photon conversion becomes strongly suppressed. Consequently, contributions to the conversion probability from regions with $r \gtrsim 40~\mathrm{kpc}$ are negligible, justifying the truncation of the integration region at this scale. We emphasize that $R$ should be interpreted as an effective oscillation length, capturing the dominant contribution to the conversion process rather than a sharply defined physical boundary.
This expression makes explicit dependence of the photon flux on both the GW strain amplitude and the properties of the astrophysical environment.

 This photon flux is expected to introduce distortions in the overall emission spectrum originating from
the central region of the M87 galaxy. The observed spectral data from M87 have been published by the Multi-
Wavelength Science Working Group Collaboration~\cite{EventHorizonTelescope:2021dvx}.
A detailed discussion of this multi-wavelength dataset is
presented in the following section.

The above relations demonstrate that magnetized environments around SMBHs can act as indirect probes of stochastic gravitational waves, converting a fraction of the GW energy into electromagnetic radiation. By comparing the predicted photon flux from graviton--photon conversion with observed electromagnetic spectra, one can place conservative upper bounds on the stochastic GW background amplitude and spectral density parameter.

\begin{figure*}
    \centering
    \includegraphics[width=0.7\linewidth]{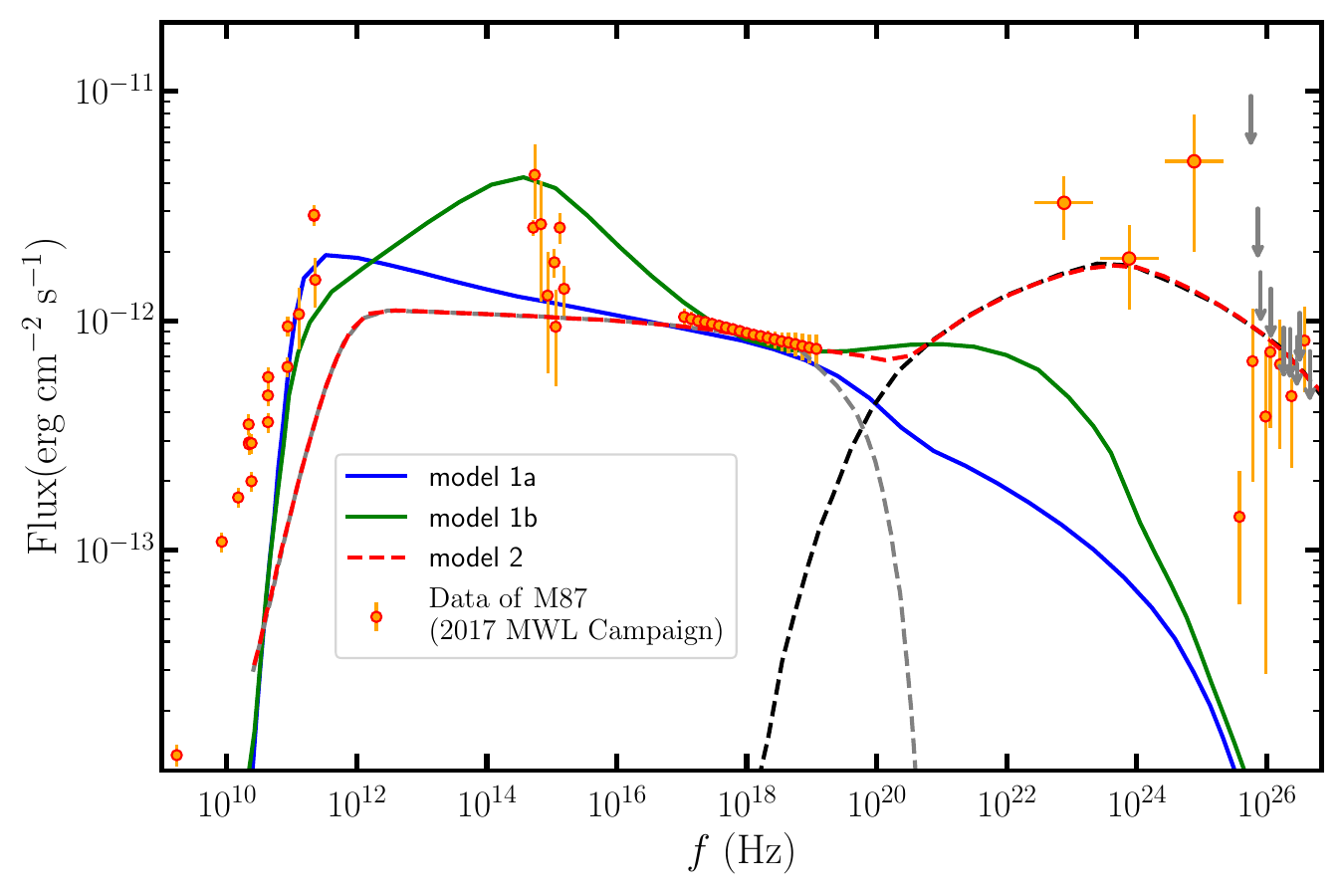}
    \caption{Multi wavelength broadband spectral energy distribution (SED) of M87 observed during the EHT campaign in April 2017. Flux measurements from various instruments are shown in orange circles with error bars, while upper limits are indicated by downward arrows. SED fits focusing on the EHT data are shown by the blue and green curves, corresponding to models 1a and 1b, respectively. The fit emphasizing the higher-energy data (model 2) is shown by the red dashed curve. The figure is adopted from \cite{EventHorizonTelescope:2021dvx}.}
    \label{fig:Astrophy_models}
\end{figure*}

\section{The Multi wavelength electromagnetic spectrum of M87 }\label{MWL_data_spectrum}
The electromagnetic spectrum employed in this study is sourced from the multi-wavelength campaign of the M87 galaxy conducted in 2017~\cite{EventHorizonTelescope:2021dvx}.
This campaign compiled observational data across a wide range of frequencies from radio to very high energy gamma rays using a coordinated effort involving numerous observatories. 
These observational data are well interpreted by established astrophysical emission models, which are also detailed in Ref.~\cite{EventHorizonTelescope:2021dvx}. These models, considered as the standard astrophysical background, leave limited room for additional signals such as those potentially arising from graviton-photon conversion.

Fig.~\ref{fig:Astrophy_models} illustrates the single–zone SED models used to characterize the 
compact emission region of M87 and to establish the astrophysical photon 
background relevant to our analysis.  
Fig.~\ref{fig:Astrophy_models} shows the SED fits constructed to reproduce the EHT-scale emission, 
with models~1a and~1b shown in blue and green, respectively.  
Both variants are tuned to match the radio-to-mm flux of the EHT core, where 
synchrotron emission dominates, and each model generates an associated 
Synchrotron Self Compton (SSC) component that appears in the $\gamma$-ray band.  
As evident in the figure, these SSC features fall below the observed $\gamma$-ray 
fluxes, demonstrating that the EHT-resolved region alone cannot account for the 
high-energy emission.  
Fig.~\ref{fig:Astrophy_models} also presents the high-energy–oriented model~2 (dashed red curve), which is  constructed using a larger emission region and is fitted primarily to the optical, 
X-ray, and $\gamma$-ray data.  
This model captures the higher-energy spectrum more effectively, though it no 
longer satisfies the EHT-scale constraints at lower frequencies.

For the purposes of this work, we adopt both the EHT-oriented models (1a/1b) and 
the high-energy–oriented model~2 as representative descriptions of the 
astrophysical photon emissions from M87.  
These models provide physically motivated backgrounds onto which we can 
superimpose the photon signal produced through graviton–photon conversion.  
Using SEDs constrained by VLBI and multi-wavelength observations ensures that 
our predicted conversion signal is evaluated against realistic emission 
environments corresponding to both compact and more extended regions of the 
central region in M87.  
Together, these models allow us to explore the graviton-induced photon spectra 
under well-defined astrophysical conditions spanning the radio through $\gamma$-ray 
bands.

\begin{figure*}
    \centering
    \includegraphics[width=8.5cm, height=6.5cm]{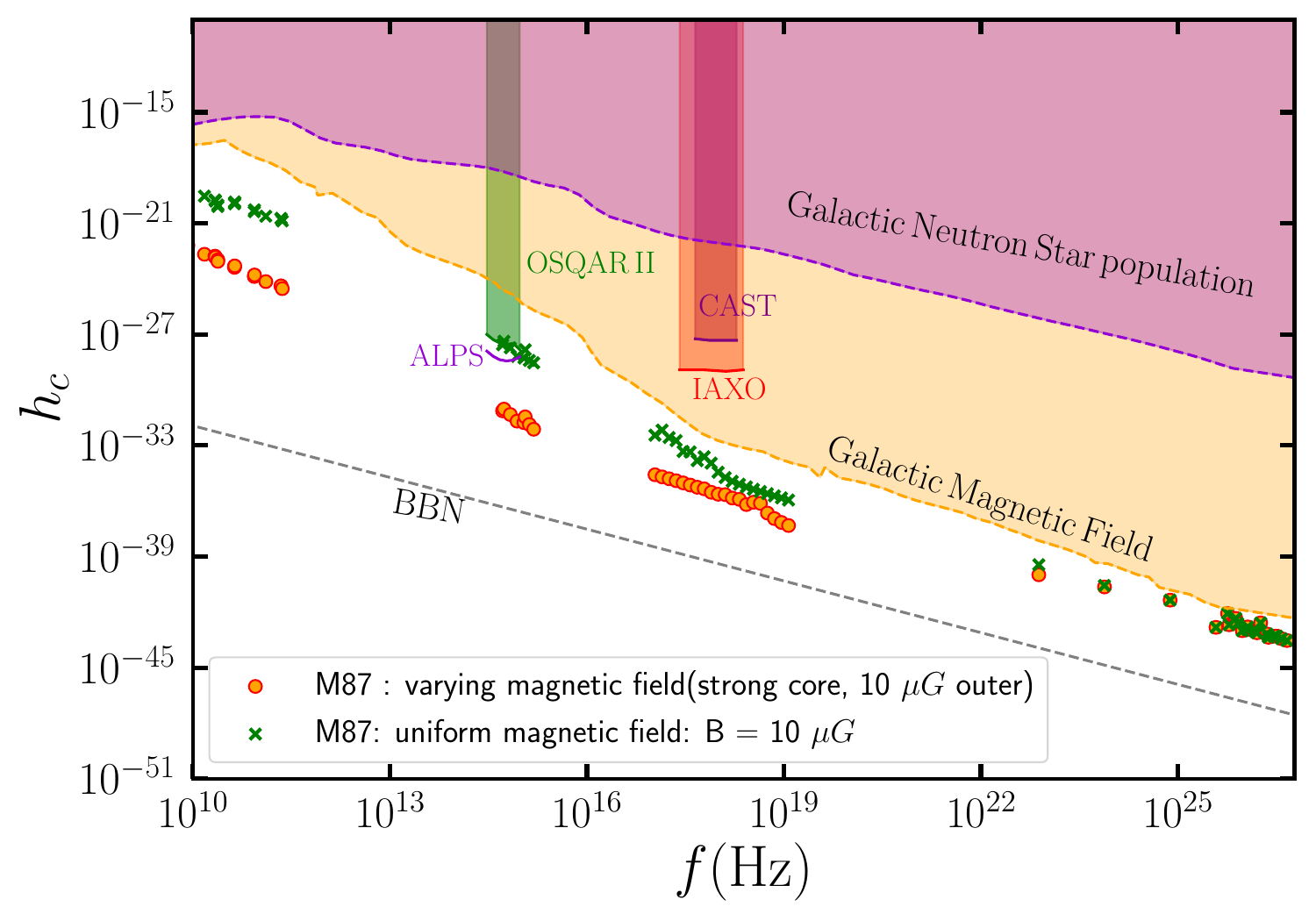}
    \includegraphics[width=8.5cm, height=6.5cm]{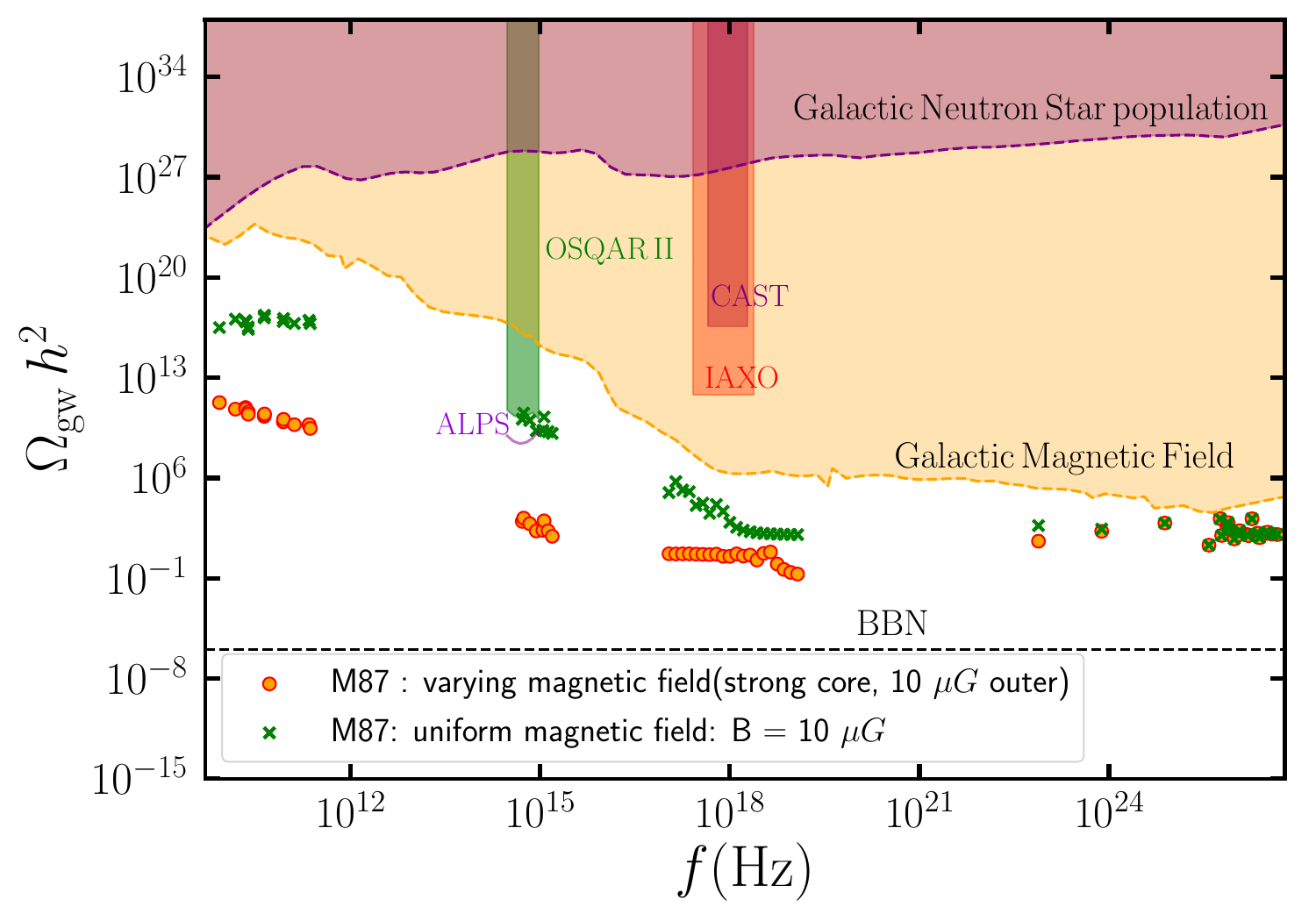}
    \caption{
Conservative constraints derived from the M87 analysis assuming a propagation distance
of $z = 40\,\mathrm{kpc}$. 
\textit{Left panel}: Upper limits on the characteristic strain amplitude $h_c$ as a
function of frequency for two magnetic-field configurations. The golden circles
correspond to a spatially varying magnetic-field profile, with a strong field near the
supermassive black hole and an average field strength of $10\,\mu\mathrm{G}$ in the outer
regions of the galaxy, while the green crosses assume a uniform magnetic field of
$10\,\mu\mathrm{G}$ across the entire distance scale.
\textit{Right panel}: Corresponding constraints on the gravitational-wave energy density
$\Omega_{\mathrm{gw}} h^2$ as a function of frequency. Existing bounds from the literature,
obtained using high-frequency gravitational-wave searches, are also shown for comparison. 
}
    \label{fig:conservative_constraints}
\end{figure*}

\begin{figure*}
    \centering
    \includegraphics[width=8.5cm, height=6.5cm]{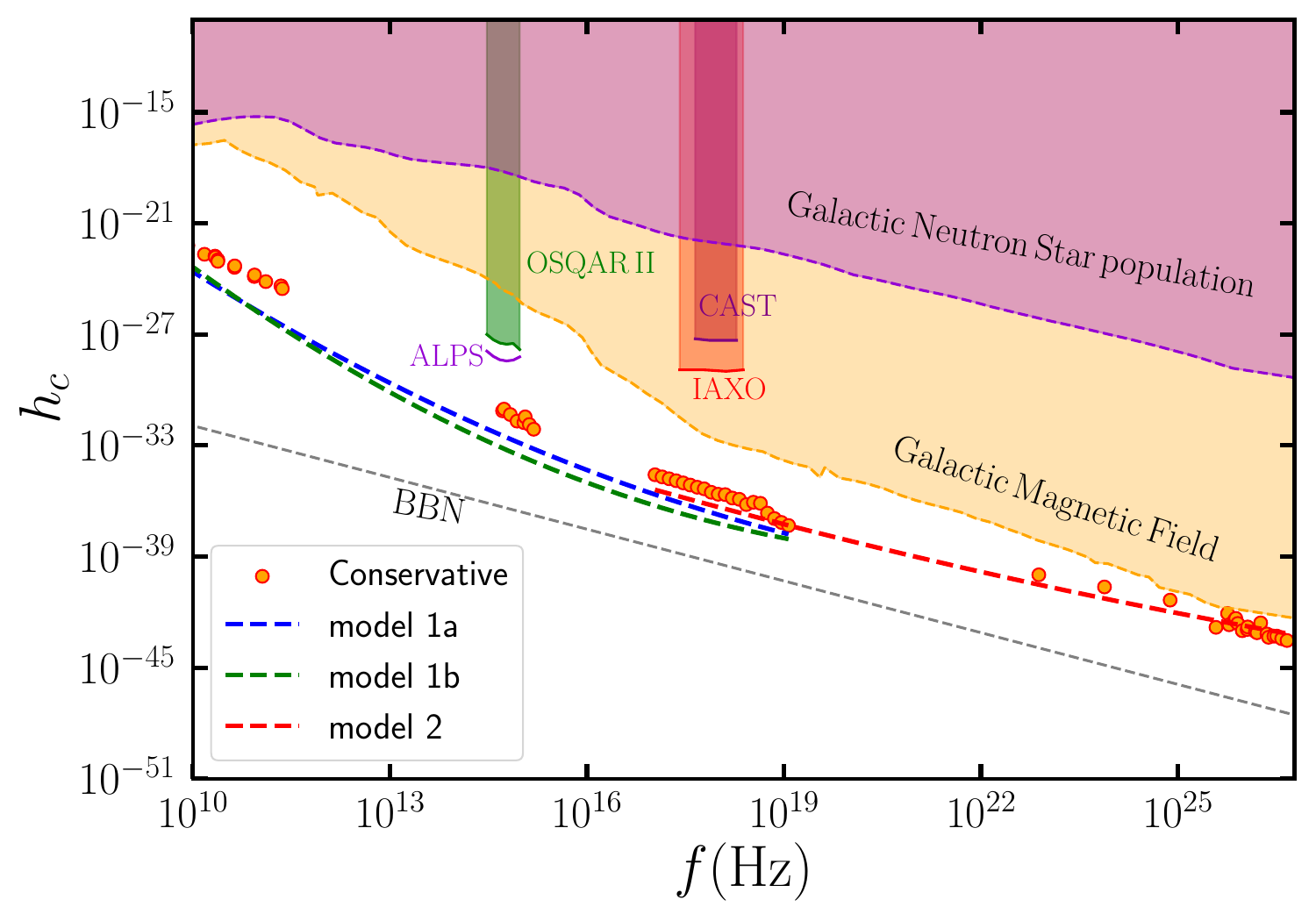}
    \includegraphics[width=8.5cm, height=6.5cm]{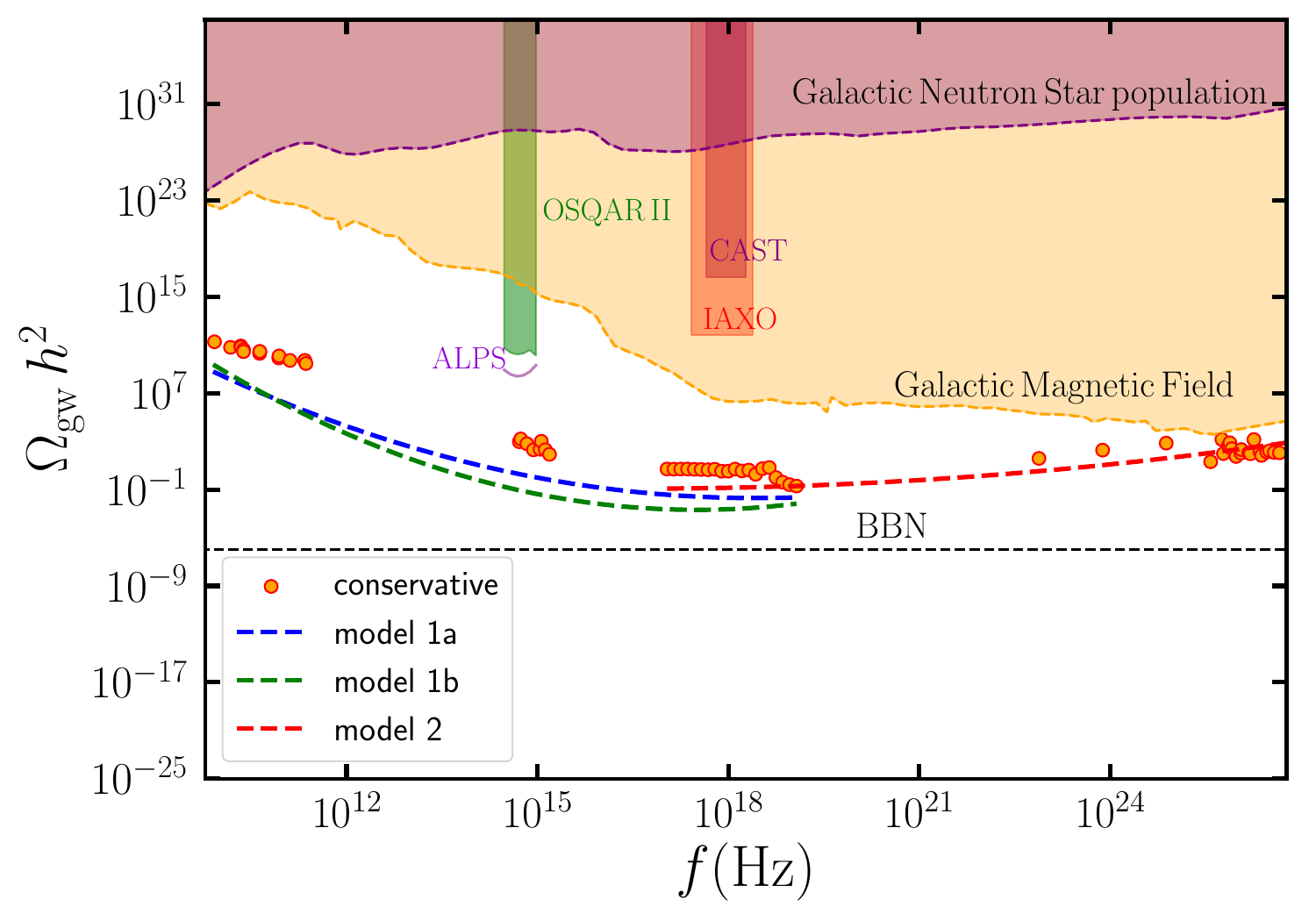}
    \caption{
Impact of different astrophysical models on the constraints obtained from
the M87 analysis.
\textit{Left panel}: Comparison of conservative limits on the characteristic strain
amplitude $h_c$ for various astrophysical models, including Model~1a, Model~1b,
and Model~2, as described in Fig.~\ref{fig:Astrophy_models}. Incorporating these
astrophysical scenarios lead to an overall strengthening of the bounds across the full
frequency range considered.
\textit{Right panel}: Corresponding constraints on the gravitational-wave energy density
$\Omega_{\mathrm{gw}} h^2$ for the same set of astrophysical models.
}
    \label{fig:constraints_w_backgrounds}
\end{figure*}

\section{Constraints on stochastic gravitational waves}

The conversion of gravitons into photons from a background graviton abundance can lead to an observable photon flux in the frequency range
\(10^{10}\,\mathrm{Hz} \lesssim f \lesssim 10^{27}\,\mathrm{Hz}\).
In the previous section, we reviewed astrophysical emission models that fit the observed broadband SED of the M87 galaxy. In particular, the EHT-oriented models (models~1a and~1b) provide an accurate description of the low-frequency emission, while a high-energy–oriented model (model~2) accounts for the observed spectrum at higher photon energies, as discussed in detail in the observational data analysis~\cite{EventHorizonTelescope:2021dvx}.

Graviton--photon conversion in the magnetized environment of M87 contributes an additional, nonstandard component to the photon flux. When astrophysical background models are specified, constraints on the graviton-induced signal can be strengthened by requiring that this additional contribution does not exceed the residual difference between the observed flux and the model-predicted background flux in each frequency bin. Accordingly, we impose
\begin{equation}
\label{constraint_bkgsub}
\Phi_{h\rightarrow\gamma}\Big|_{f=f_i}
\;\lesssim\;
\left|
\Phi_{\gamma,i}^{\rm obs}
-
\Phi_{\gamma,i}^{\rm bkg}
\right|,
\end{equation}
where \(\Phi_{\gamma,i}^{\rm obs}\) denotes the observed photon flux in the \(i\)-th frequency bin and \(\Phi_{\gamma,i}^{\rm bkg}\) is the corresponding flux predicted by the astrophysical emission model under consideration. In Fig.~\ref{fig:Astrophy_models}, the observed flux is shown in units of $\rm erg\, cm^{-2}\, s^{-1}$, where the quantity plotted is frequency times flux. The theoretical graviton-induced photon flux \(\Phi_{h\rightarrow\gamma}\) is computed as described in Eq.~\eqref{grav_induced_phflux}.

To quantify the resulting constraints more rigorously, we perform a binned \(\chi^2\) analysis over the relevant frequency range. For a given value of the unknown parameter, taken here to be the characteristic strain amplitude \(h_c\), we define
\begin{equation}
\chi^2(h_c)
=
\sum_i
\frac{
\left[
\Phi_{\gamma,i}^{\rm obs}
-
\left(
\Phi_{\gamma,i}^{\rm bkg}
+
\Phi_{h\rightarrow\gamma}(h_c)
\right)
\right]^2
}{
\sigma_i^2
},
\end{equation}
where \(\sigma_i\) denotes the experimental uncertainty in the \(i\)-th frequency bin. Upper limits on \(h_c\) are then obtained by requiring
\begin{equation}
\Delta\chi^2 \equiv \chi^2(h_c) - \chi^2_{\rm min} = 2.71,
\end{equation}
corresponding to a one-parameter constraint at the 95\% confidence level.

For comparison, we also present a conservative bound obtained without assuming any specific astrophysical background model. In this case, the graviton-induced photon flux is required to remain below the observed flux itself in each frequency bin,
\begin{equation}
\label{constraint_conservative}
\Phi_{h\rightarrow\gamma}\Big|_{f=f_i}
\;\lesssim\;
\Phi_{\gamma,i}^{\rm obs}.
\end{equation}

\begin{figure*}
    \centering
    \includegraphics[width=0.65\linewidth]{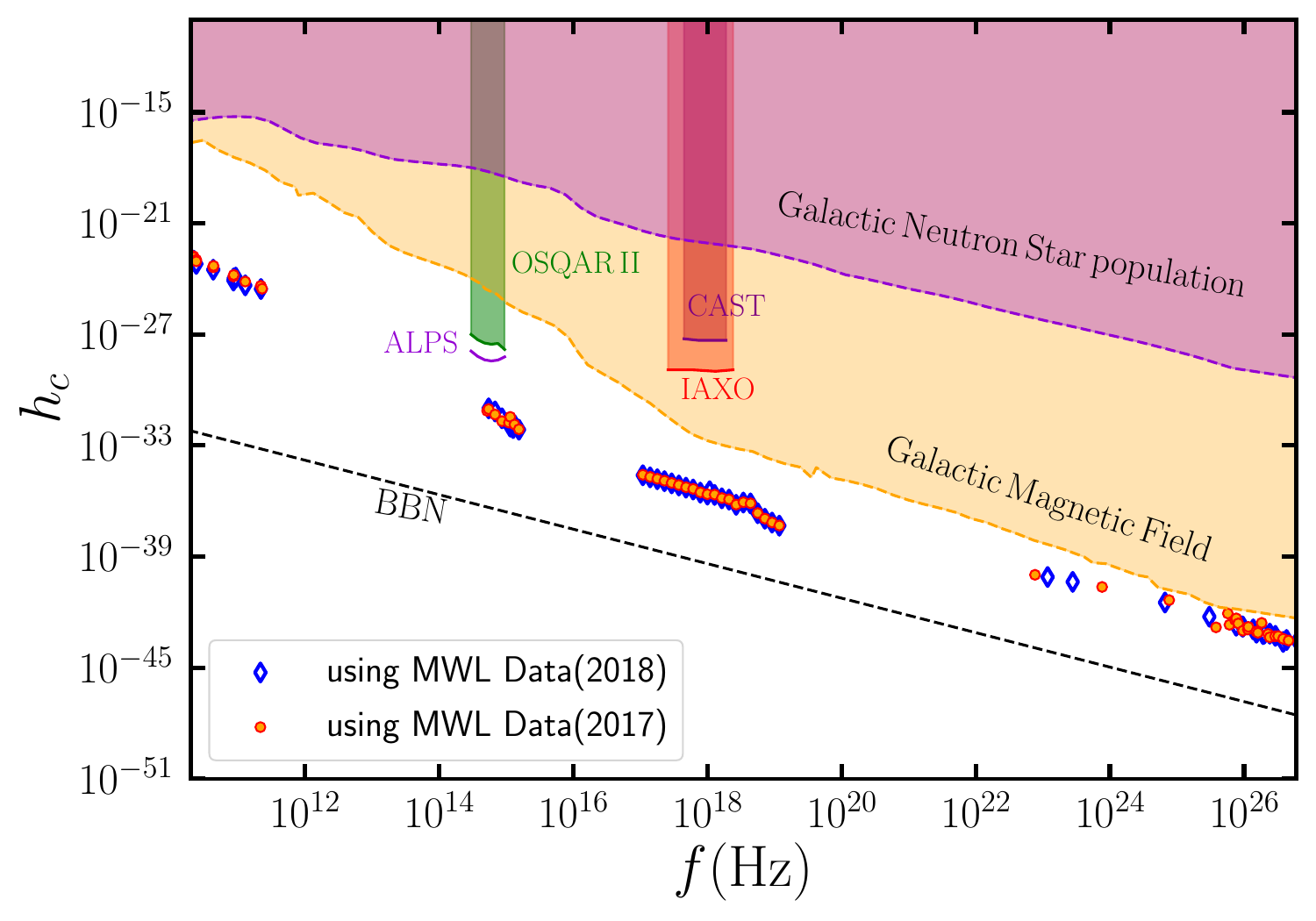}
    \caption{Comparison of the conservative constraints derived from two independent multi-wavelength observation campaigns of M87, represented by the golden circles and blue diamonds.}
    \label{fig:compare_dataset}
\end{figure*}

In the left panel of Fig.~\ref{fig:conservative_constraints}, we present conservative constraints on the strain amplitude
$h_c$ derived from the analysis of M87, assuming a characteristic propagation distance
of $z = 40\,\mathrm{kpc}$. We show results for two magnetic-field configurations.
The first corresponds to a spatially varying magnetic field profile, characterized by a
strong field in the immediate vicinity of the supermassive black hole and an average
field strength of $10\,\mu\mathrm{G}$ in the outer regions of the galaxy. The corresponding
constraints are indicated by golden circles for different observed frequencies, as
illustrated in Fig.~\ref{fig:conservative_constraints}. The second configuration assumes a uniform
magnetic field of $10\,\mu\mathrm{G}$ across the entire distance scale, with the resulting
limits shown by green crosses. The spatially varying field yields stronger constraints at frequencies $\lesssim 10^{15}$ Hz because the enhanced magnetic field in the inner region boosts the conversion probability at these energies. At higher frequencies, the conversion occurs predominantly in the outer region where both profiles converge to $10 \mu$G, leading to similar constraints.

The right panel of Fig.~\ref{fig:conservative_constraints} displays the corresponding constraints on the gravitational-wave
spectral energy density, $\Omega_{\mathrm{gw}} h^2$, as a function of frequency. For comparison, we also include existing bounds relevant to astrophysical scenarios from the literature~\cite{Aggarwal:2025noe}.
The datasets corresponding to these bounds—such as those derived from galactic magnetic fields, galactic neutron star populations, and detector sensitivities of OSQAR, ALPSII, CAST, and IAXO are extracted using the high-frequency gravitational-wave compilation tool~\cite{HFGW_web}.
As is evident from the figure, our
constraints are stronger than those obtained in earlier studies employing a similar
methodology based on the Milky Way magnetic field.

 In the left panel of Fig.~\ref{fig:constraints_w_backgrounds}, we further compare these conservative constraints with those
derived under different astrophysical background models for the emitted photons, namely Model~1a, Model~1b,
and Model~2, as described in Fig.~\ref{fig:Astrophy_models}. In this case, we incorporate the spatially varying magnetic field in the probability calculations, as shown by the red curve in Fig.~\ref{fig:num_prob}. Incorporating these
astrophysical scenarios lead to a significant improvement in the bounds, particularly when the background contribution is taken into account in the graviton-mediated photon
search within the emission spectrum of M87 across the full frequency range considered.
The corresponding constraints on $\Omega_{\mathrm{gw}} h^2$ are shown in the right panel
of Fig.~\ref{fig:constraints_w_backgrounds}.

 As expected, the background-subtracted analysis based on realistic emission models yields stronger constraints than this conservative approach, while remaining robust against uncertainties in the modeling of the emitted photons from M87. This improvement is particularly visible in the frequency range $10^{10} - 10^{19}$ Hz (radio to X-ray bands, where the models closely track the observations) on comparing Figs.~\ref{fig:conservative_constraints} and~\ref{fig:constraints_w_backgrounds}.

In addition, we present the resulting constraints derived using the dataset from the 2018 multi-wavelength (MWL 2018)
campaign of M87 in Fig.~\ref{fig:compare_dataset}. This dataset corresponds to the flaring episode observed in M87 during the 2018 campaign~\cite{EventHorizonTelescope-Multi-wavelengthscienceworkinggroup:2024xhy}.
The constraints obtained from this dataset are shown as blue diamonds, adopting the same conservative treatment as for
the 2017 MWL campaign data analysis, which is represented by orange circles, as shown in Fig.~\ref{fig:conservative_constraints} and Fig.~\ref{fig:constraints_w_backgrounds}.

Our results indicate that the 2018 MWL dataset does not lead to any significant improvement or qualitative change in
the derived constraints compared to those obtained from the earlier, relatively stable 2017 dataset. This suggests that
the graviton-induced photon emission originating from the magnetospheric region of M87 is largely insensitive to transient
flaring activity and is instead dominated by the underlying, steady-state emission properties of the system.

The stronger constraints obtained in our analysis follow directly from the dependence of the predicted electromagnetic flux on the graviton-to-photon conversion probability and on the source geometry. In the Milky Way scenario of Ref.~\cite{Lella:2024dus}, the conversion occurs in the Galactic magnetic field, and the relevant photon flux is the Cosmic Photon Background (CPB) taken from Ref.~\cite{Hill:2018trh} (see also Fig.~5 of Ref.~\cite{Lella:2024dus}), which at $f \sim 10^{18} {\rm Hz}$ is of order $\Phi_{CPB} = 10^{-7}\, {\rm erg \, cm^{-2} \, s^{-1}\,sr^{-1}}$. Because the Galactic magnetic field is comparatively weak, the corresponding conversion probability is small, $P_{h\to\gamma}^{MW} \approx 7.8\times10^{-17}$. By contrast, the magnetic field in the central region of M87 is substantially stronger, yielding a conversion probability of  $P_{h\to\gamma}^{M87} \approx 1.2\times10^{-13}$ even in the conservative constant-field approximation (see Fig.~\ref{fig:num_prob}) — an enhancement of roughly $1600$ times that of the Milky Way case.\\
To compare the constraints from the two scenarios, we scale the CPB flux by $10^{-7}
\,{\rm erg}\,{\rm cm}^{-2}\,{\rm s}^{-1}\,{\rm sr}^{-1}
\left(\frac{R}{d}\right)^2.$ In this case, the flux would be suppressed by the geometric factor ($(R/d)^2$). For M87, taking $R=40\,{\rm kpc}$ and $d\simeq16.8\,{\rm Mpc}$, one finds $R^2/d^2 \simeq 6\times10^{-6}$ and the corresponding equivalent flux from CPB is therefore $6\times10^{-6}\times 10^{-7}\,{\rm erg\,cm^{-2}\,s^{-1}} sr^{-1}$ which is about $6\times 10^{-13}\,{\rm erg\,cm^{-2}\,s^{-1}} sr^{-1}$. While the observed flux from M87 for $f \gtrsim 10^{18}\,{\rm Hz}$
is of order $10^{-12}$ (see Fig.~\ref{fig:Astrophy_models}).

 Comparing the constraints from the two scenarios, and using Eq.~(\ref{grav_induced_phflux}), the ratio of strain amplitudes is

\begin{equation}
\frac{\left(h_c^{\rm M87}\right)^2}{\left(h_c^{\rm MW}\right)^2} =\frac{P_{h\to\gamma}^{\rm MW}}{P_{h\to\gamma}^{\rm M87}} \frac{10^{-12}}{6\times10^{-13}}\Bigg|_{f=10^{18}\,Hz}
\end{equation}

\begin{equation}
\frac{h_c^{\rm M87}}{h_c^{\rm MW}} =\sqrt{\frac{7.8\times 10^{-17}}{1.2\times 10^{-13}} \frac{10^{-12}}{6\times10^{-13}}} \sim 10^{-2},
\end{equation} 

implying that the bound obtained from M87 is approximately two orders of magnitude stronger than that derived from the Milky Way and the CPB fluxes considered in Ref.~\cite{Lella:2024dus}.


\section{Conclusion}
In this study, we examine the detection potential for high-frequency gravitons to convert into photons within the magnetic field environment of the M87 galaxy. For this purpose, we utilize the known magnetic field profile of M87, which notably incorporates the influence of its central SMBH.

The measured electromagnetic spectrum from M87 spans a large frequency range, from $10^{10}\,\mathrm{Hz}$ to $10^{27}\,\mathrm{Hz}$. These broadband observations were compiled and made available by the MWL Working Group during their 2017 observational campaign. Utilizing these data, we derive stringent constraints on the characteristic strain $h_c$ and corresponding energy density $\Omega_{\mathrm{gw}} h^2$ of the stochastic gravitational wave background.

As illustrated in Fig.~\ref{fig:conservative_constraints} and Fig.~\ref{fig:constraints_w_backgrounds}, our derived constraints on $h_c$—shown as golden circles for the conservative case and dashed lines (blue, green and red) when astrophysical backgrounds are incorporated—are approximately 1--5 orders ($\sim \mathcal{O}(1)$ for $\gamma-$rays, $\sim \mathcal{O}(4)$ in X-rays and $\sim \mathcal{O}(5)$ for radio waves) of magnitude more stringent than those reported in Ref.~\cite{Lella:2024dus} (yellow dashed line) over the same frequency range. Their analysis is based on the Milky Way’s magnetic field and its associated cosmic photon background. 
We also compare our bounds with those inferred from graviton–photon conversions in the magnetospheres of galactic neutron stars (dashed purple line). In the right panels of both figures, the constraints on the energy density parameter $\Omega_{\mathrm{gw}} h^2$ are shown.

Several experimental efforts aim to detect high-frequency GWs via graviton-to-photon conversion process. A representative selection of the corresponding experimental constraints is shown in Fig.~\ref{fig:conservative_constraints} and Fig.~\ref{fig:constraints_w_backgrounds}. For instance, the OSQARII experiment excludes strain amplitudes of $h_c \gtrsim 10^{-26}$ over the frequency range $2.7 \times 10^{14}\,\mathrm{Hz} \lesssim f \lesssim 1.4 \times 10^{15}\,\mathrm{Hz}$. In contrast, the CAST experiment sets substantially stronger limits, improving by roughly five orders of magnitude in $h_c$ across the range $5 \times 10^{18}\,\mathrm{Hz} \lesssim f \lesssim 1.2 \times 10^{19}\,\mathrm{Hz}$. Looking ahead, the IAXO experiment is expected to probe strain amplitudes as small as $h_c \sim 10^{-29}$ in the frequency band $f \sim 10^{17}$--$10^{18}\,\mathrm{Hz}$~\cite{Ejlli:2019bqj}. The ALPS experiment operates in a frequency range similar to that of OSQAR~II but can constrain strain amplitudes down to $h_c \sim 10^{-29}$.

In addition to these laboratory constraints, we also present the bound from Big Bang Nucleosynthesis (BBN), indicated by the black dashed curve. This limit is derived by requiring that the energy density in gravitational waves, $\rho_{\rm gw}$, does not exceed the contribution allowed by the effective number of relativistic species, $N_{\rm eff}$~\cite{Pagano:2015hma}

\begin{equation}
    \frac{\rho_{\rm gw}}{\rho_{\gamma}} \lesssim \frac{7}{8} \left(\frac{4}{11}\right)^{4/3} N_{\rm eff}
\end{equation}

where $\rho_{\gamma}$ is the present-day photon energy density, and current limits from CMB and BBN observations suggest $\Delta N_{\rm eff}(\equiv N_{\rm eff}-3.046) \lesssim 0.3$~\cite{Planck:2018vyg,Cyburt:2015mya}. As seen in the plot, this cosmological bound is significantly more stringent than astrophysical ones. However, it does not apply to gravitational wave sources active after the epoch of CMB decoupling, such as the scenario under consideration in this work.

Despite current limitations, the sensitivity of telescopes and experiments targeting cosmic electromagnetic radiation at high frequencies is approaching the constraints set by cosmological observations. As a result, future advancements in X-ray and gamma-ray astronomy may present promising opportunities for detecting a GWB at very high frequencies. Notably, in the X-ray regime, the upcoming Athena X-ray observatory is anticipated to significantly lower the detection threshold for astrophysical sources~\cite{Cucchetti:2018izp}. This improvement will not only refine measurements of unresolved diffuse emission but also enhance our understanding of the astrophysical sources that contribute to it, thereby increasing the potential to identify unconventional or exotic signals.

Similarly, future missions targeting the MeV energy range, such as the proposed
COSI telescope~\cite{cosi_web,Tomsick:2023aue},
are expected to provide valuable insights. At higher energies, in the GeV to TeV
bands, observatories such as the High Energy cosmic-Radiation Detection (HERD)
facility~\cite{herd_web,Kyratzis:2020trm,Gargano:2022wih} and the Cherenkov
Telescope Array~\cite{CTAConsortium:2017dvg}, through its forthcoming extragalactic
survey, are poised to significantly advance our understanding of the composition
of the gamma-ray background. In addition to the improved sensitivity expected from future observations, further progress in probing graviton--photon conversion in astrophysical environments such as M87 will crucially depend on even more precise characterization of the underlying magnetic field structure, realistic plasma density profiles, and the geometry of the emission region. In particular, mapping the radial evolution of the magnetic field geometry from the black hole to kiloparsec scales—through future high-sensitivity VLBI~\cite{Johnson:2023ynn,2024evn..conf..159R,Johnson:2024ttr} and multiwavelength polarimetric observations~\cite{ALMA:2025wvr}—will be essential for reducing systematic uncertainties in graviton–photon conversion models. Incorporating these astrophysical ingredients in a self-consistent manner is essential for reducing theoretical uncertainties and enhancing the robustness of predicted signals. Together, these developments offer a promising pathway toward exploring stochastic gravitational wave backgrounds at frequencies beyond the reach of conventional detectors.

\section{Acknowledgement}
PS gratefully acknowledges financial support from the University Grants Commission, Government of India, in the form of a Senior Research Fellowship.
\section{Data Availability}
The data are available from the authors upon reasonable request.

\appendix
\section{Analytical form of graviton to photon conversion probability}\label{appenA}
To make progress analytically, we consider a simplified setup. Suppose the magnetic field in the central region of the M87 galaxy is spatially uniform and oriented along the $z$-axis. In this context, the photon polarization mode parallel to the magnetic field, denoted as $\parallel$, aligns with $B_T$ (i.e. $\cos\phi = 1$). This assumption allows the original $4\times4$ system of differential Eqs. to decouple into two separate $2\times2$ systems, each of which can be treated analytically. Consequently, the polarization modes $\left\{ A_{\parallel},h_{\times} \right\}$ and $\left\{ A_{\perp},h_{+} \right\}$ evolve independently. Accordingly, we have

\begin{equation}
    \Bigg[i\frac{d}{dz}+\omega\Bigg]
    \begin{pmatrix}
        A_{\perp}\\
        h_{+}
    \end{pmatrix}
    =
    \begin{pmatrix}
        \Delta_{\perp} &  \Delta_{g\gamma}\\
         \Delta_{g\gamma} & 0
    \end{pmatrix}
    \begin{pmatrix}
        A_{\perp}\\
        h_{+}
    \end{pmatrix}
\end{equation}

An analogous equation holds for the other pair, $\left\{ A_{\parallel},h_{\times} \right\}$. This system can be solved by diagonalizing the mixing matrix. The eigenvalues of the matrix are

\begin{equation}
     e_{1,2} = \frac{1}{2}\left[\Delta_{\perp} \pm \sqrt{\Delta_{\perp}^2 + 4\Delta_{g\gamma}^2}\right] \quad ; \quad \mathcal{D} = \begin{pmatrix}
         e_{1} & 0 \\
         0 & e_{2}
     \end{pmatrix}
\end{equation}

Letting $\psi=(A_{\perp}, h_{+})^T$ denote the state vector, we apply a unitary transformation $\psi \rightarrow \psi' = \mathcal{U} \psi$ with

\begin{equation}
     \mathcal{U}=
     \begin{pmatrix}
         \cos \theta & \sin \theta \\
         -\sin \theta & \cos \theta
     \end{pmatrix} \quad ; \quad \tan 2\theta = \frac{2\Delta_{g\gamma}}{\Delta_{\perp}}
\end{equation}

This rotation diagonalizes the interaction matrix, i.e.,
\[
\mathcal{U}
 \begin{pmatrix}
     \Delta_{\perp} &  \Delta_{g\gamma}\\
     \Delta_{g\gamma} & 0
 \end{pmatrix}\mathcal{U}^{-1} = \mathcal{D}, \quad \text{with } \mathcal{U}^{-1} = \mathcal{U}^{T}.
\]

In the rotated frame, the evolution equation simplifies to

\begin{equation}
     \Bigg[i\frac{d}{dz}+\omega\Bigg]\psi' = \mathcal{D}\psi'
\end{equation}

with the general solution:

\begin{equation}
     \psi'= e^{-i(\mathcal{D}-\omega)z} \psi'_{ini}(z_{ini})
\end{equation}

Transforming back to the original basis yields

\begin{equation}
     \psi = e^{i\omega z} \mathcal{U}^{T} e^{-i\mathcal{D}z}\mathcal{U} \psi_{ini}(z_{ini}) 
\end{equation}

Assuming the initial condition corresponds to a purely graviton state, $\psi_{ini}(z_{ini}) = (0,1)^{T}$, the evolution operator $\mathcal{K} = \mathcal{U}^{T} e^{-i\mathcal{D}z} \mathcal{U}$ encapsulates the transition amplitudes
\begin{widetext}
\begin{equation}
\begin{aligned}
    \mathcal{K} &=
    \begin{pmatrix}
        \mathcal{K}_{11} &  \mathcal{K}_{12}\\
        \mathcal{K}_{21} &  \mathcal{K}_{22}
    \end{pmatrix} 
    &=
    \begin{pmatrix}
        [\cos^2\theta \, e^{-ize_1} + \sin^2\theta \, e^{-ize_2}] 
        & [\sin\theta \cos\theta \left(e^{-ize_1} - e^{-ize_2}\right)]\\
        [\sin\theta \cos\theta \left(e^{-ize_1} - e^{-ize_2}\right)] 
        & [\cos^2\theta \, e^{-ize_2} + \sin^2\theta \, e^{-ize_1}]
    \end{pmatrix}
\end{aligned}
\end{equation}
\begin{figure*}
    \centering
    \includegraphics[width=8cm, height=5.8cm]{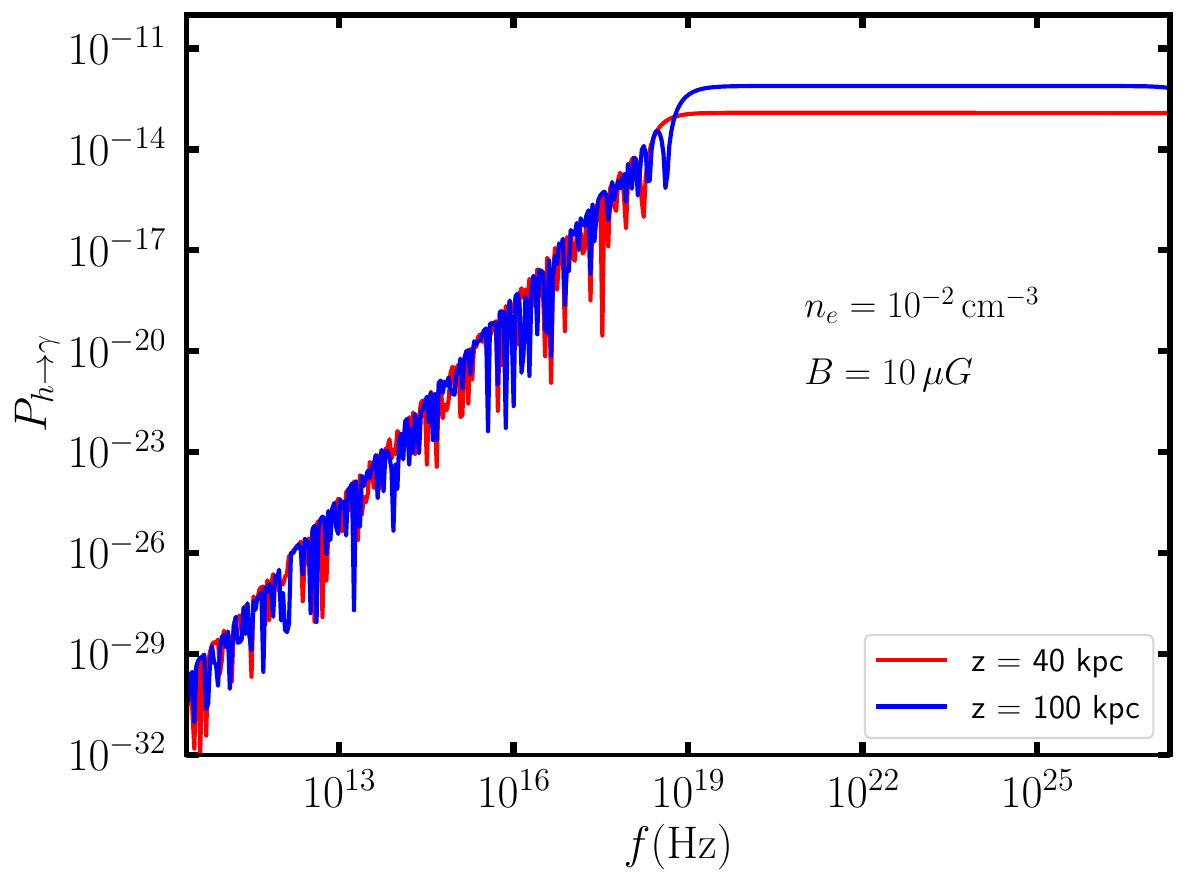}
    \includegraphics[width=8cm, height=5.8cm]{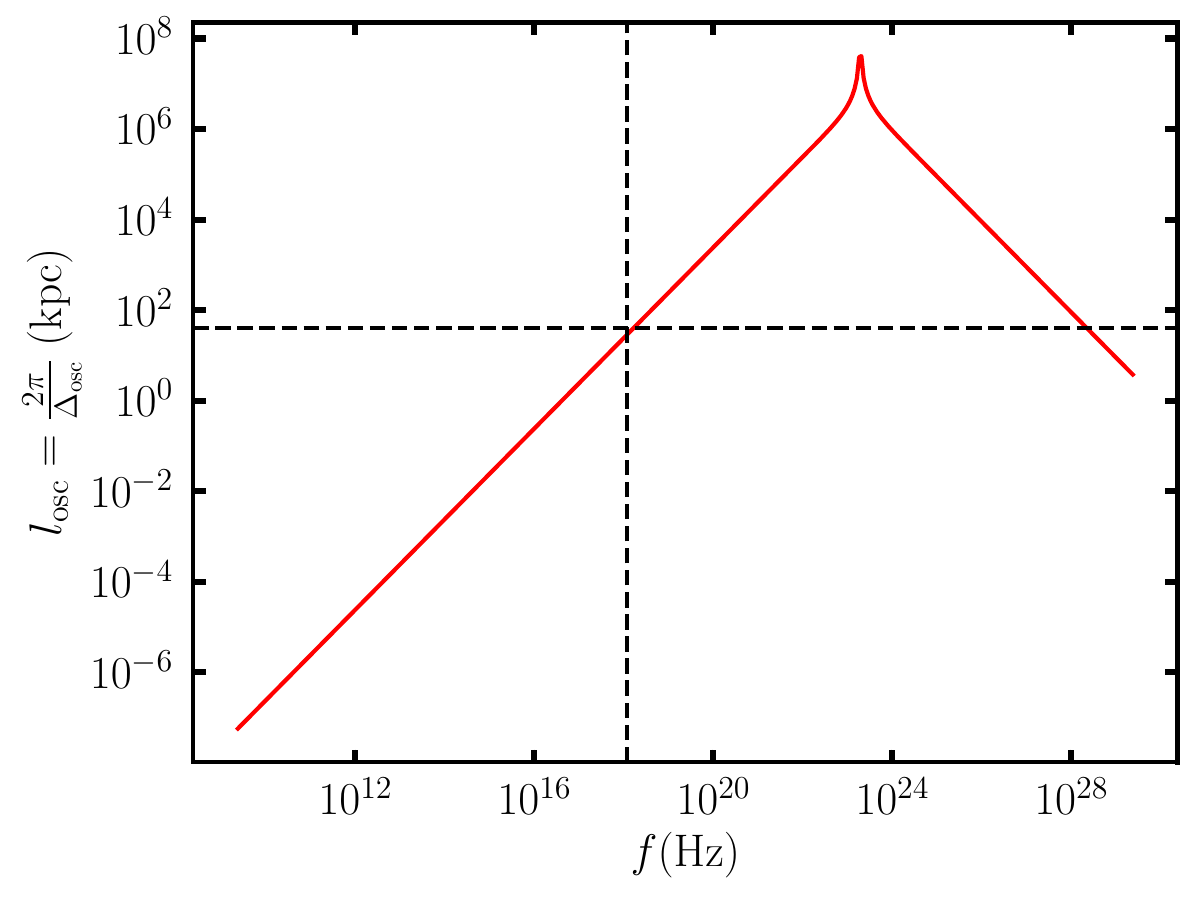}
    \caption{
Left panel: Graviton--photon conversion probability $P_{g\gamma}$ as a function of the frequency $f$, computed using the analytical expression in Eq.~\eqref{Eq:prob_analytical}.
The analytical result (i.e. red curve for $z=40$ kpc) is in excellent agreement with the corresponding numerical solution shown in Fig.~\ref{fig:num_prob} (green curve), thereby validating the numerical procedure.
Right panel: Oscillation length $\ell_{\mathrm{osc}} = 2\pi/\Delta_{\mathrm{osc}}$ as a function of $f$ for the same choice of physical parameters. At low frequencies, the behavior is dominated by the plasma contribution $\Delta_{\mathrm{pl}} \propto f^{-1}$, leading to a linear increase of the oscillation length with frequency. In the frequency range $f \sim 10^{18}$--$10^{28}\,\mathrm{Hz}$, the condition $\ell_{\mathrm{osc}} \gg z$ is satisfied, placing the system in the coherent conversion regime, where $P_{h\to\gamma}\propto z^2$.
}
    \label{fig:analytical_prob_osclength}
\end{figure*}
\end{widetext}
Given the initial state, the probability for graviton-to-photon conversion is given by the squared modulus of the off-diagonal element $|\mathcal{K}_{12}|^2$

\begin{widetext}
\begin{equation}\label{Eq:prob_analytical}
    P(z) = | \langle h_{+,ini} | A_{\perp}(z) \rangle |^2 = |\sin \theta \cos \theta (e^{-ize_{1}} - e^{-ize_{2}})|^2 = \frac{4\Delta_{g\gamma}^2}{\Delta_{osc}^2} \sin^2\left( \frac{\Delta_{osc} z}{2} \right)
\end{equation}   
\end{widetext}
where
\begin{equation}
\Delta_{\text{osc}} =
\sqrt{\Delta_{\perp}^2 + 4\,\Delta_{g\gamma}^2}\,,
\label{eq:deltaosc}
\end{equation}
The expression for the oscillation length is defined as
\begin{equation}
\ell_{\text{osc}} = \frac{2\pi}{\Delta_{\text{osc}}}
\end{equation}
An analogous expression for the conversion probability of the other polarization channel,
$\left( h_{\times} \to A_{\parallel} \right)$, can be obtained with only a minor modification.
Specifically, the form of the oscillation term $\Delta_{\mathrm{osc}}$ remains unchanged,
except that the parameter $\kappa = 2$ appearing in the previous case is replaced by
$\kappa = 7/2$ for this photon polarization.

The conversion probability, computed using the analytical expression~\ref{Eq:prob_analytical} for both polarizations, is shown in the left panel of Fig.~\ref{fig:analytical_prob_osclength}.
In this figure, we adopt an electron number density $n_e = 10^{-2}\,\text{cm}^{-3}$, consistent with the plasma density expected
in the outer regions of M87 at kiloparsec scales from the galactic center. We assume a constant magnetic field strength of $10\mu G$ over a distance scale of $z=40$ kpc.
This result (i.e. the red curve with $z=40$ kpc) is in excellent agreement with the numerically evaluated curve shown in Fig.~\ref{fig:num_prob} (green line), obtained for the same choices of electron number density,
constant magnetic field strength, and oscillation distance, thereby validating our numerical analysis.

The right panel of Fig.~\ref{fig:analytical_prob_osclength} illustrates the behavior of the oscillation
length in the high-frequency regime of the photon--graviton system.
At lower frequencies, the scaling is governed by the plasma contribution,
$\Delta_{\mathrm{pl}} \propto f^{-1}$, implying that the oscillation
length grows linearly with increasing frequency.

Across the frequency interval $f \sim 10^{18} - 10^{28}\,\mathrm{Hz}$, the oscillation length satisfies $\ell_{\mathrm{osc}} \gg z$. The conversion therefore occurs coherently over the entire propagation path, leading to the characteristic $P_{h\to \gamma}\propto z^2$ scaling and an approximately constant conversion probability as a function of frequency.

Under these conditions, the mixing between
gravitons and photons becomes efficient, and the resulting conversion
probability is effectively independent of energy:
\begin{equation}\label{prob_analytical}
P_{h\to \gamma} = (\Delta_{g\gamma} z)^2
\simeq 1.2 \times 10^{-13}
\left( \frac{B_T}{10\,\mu\text{G}} \right)^2
\left( \frac{z}{40\,\text{kpc}} \right)^2 .
\end{equation}


\bibliographystyle{JHEP}
\bibliography{Final_ref.bib}

\end{document}